\documentclass[aps,pre,notitlepage,nofootinbib,superscriptaddress]{revtex4}
\usepackage{graphicx}
\usepackage{epstopdf}
\usepackage{bm}
\usepackage{amssymb}
\usepackage{psfrag}
\usepackage[final]{changes}
\usepackage{amsbsy} 
\usepackage{amsmath,amsthm,nccmath}
\usepackage{mathtools}
\usepackage{dsfont}
\usepackage{makecell}
\usepackage{lineno}
\usepackage{comment}
\usepackage{hyperref}
\usepackage{tcolorbox}
\usepackage{soul,color}
\usepackage{url}
\usepackage{bbm}
\usepackage{etoolbox}
\patchcmd{\section}
{\centering}
{\raggedright}
{}
{}
\patchcmd{\subsection}
{\centering}
{\raggedright}
{}
{}

\theoremstyle{definition}

\theoremstyle{remark}

\graphicspath{{./figs/}}
\usepackage{newfloat}
\DeclareFloatingEnvironment[name={Fig. D}]{suppfigure}
\newtheorem{thm}{Theorem}%
\newtheorem{prop}[thm]{Proposition}%

\usepackage{natbib} 
\usepackage{pifont}

\def\dd{\textrm{d}}

\def\f{\frac}

\def\p{\partial}

\def\bf{\textbf{}}

\def\Y{{\bf Y}}

\def\p{\partial}

\definecolor{red}{rgb}{0.7,0,0}
\definecolor{green}{rgb}{0,0.55,0}

\usepackage{dsfont}
\usepackage{enumerate}

\begin{document}

\title{Kinetic theories of state- and generation-dependent cell populations}
\author{Mingtao Xia}

\address{Courant Institute of Mathematical Sciences, New York University, New York, NY,
10012, USA}

\author{Tom Chou}
\address{Department of Computational Medicine, UCLA, Los Angeles, CA,
90095-1766, USA}
\address{Department of Mathematics, UCLA, Los Angeles, CA,
90095-1555, USA}





\begin{abstract}
We formulate a general, high-dimensional kinetic theory describing the
internal state (such as gene expression or protein levels) of cells in
a stochastically evolving population. The resolution of our kinetic
theory also allows one to track subpopulations associated with each
generation. Both intrinsic noise of the cell's internal attribute and
randomness in a cell's division times (demographic stochasticity) are
fundamental to the development of our model.  Based on this general
framework, we are able to marginalize the high-dimensional kinetic
PDEs in a number of different ways to derive equations that describe
the dynamics of marginalized or ``macroscopic'' quantities such as
structured population densities, moments of generation-dependent
cellular states, and moments of the total population. We also show how
nonlinear ``interaction'' terms in lower-dimensional
integrodifferential equations can arise from high-dimensional
\textit{linear} kinetic models that contain rate parameters of a cell
(birth and death rates) that depend on variables associated with other
cells, generating couplings in the dynamics. Our analysis provides a
general, more complete mathematical framework that resolves the
coevolution of cell populations and cell states. The approach may be
tailored for studying, \textit{e.g.}, gene expression in developing
tissues, or other more general particle systems which exhibit Brownian
noise in individual attributes and population-level demographic noise.
\end{abstract}


\maketitle

\section{Introduction}

Mathematical models have been formulated to describe the evolution of
populations according to a number of individual attributes such as
age, size, and/or added size since birth. Such structured population
models have various applications across diverse fields.  For example,
deterministic age-structured models that incorporate age-dependent
birth and death were developed by McKendrick and have been applied to
human populations \cite{foerster1959some,CHINA_BC}. Structured
population models have also been applied to model cell size control
\cite{JUN2015,burov2018effective}, cellular division mechanisms
\cite{DOUMIC2014}, and structured cell population models
\cite{PERTHAME2008,Metz1986}.

In a proliferating cell population, individual cell growth is
interrupted by cell division events that generate daughter
cells. Kinetic theory is a natural framework to capture the link
between individual cellular growth and division, within a
proliferating population of cells. Kinetic theories of simple
birth-death processes that track the chronological age of each cell
have been developed
\cite{chou2016hierarchical_PRE,chou2016hierarchical,Xia_2021,YATES2022} that
establish a rigorous mathematical framework to describe how individual
cell aging, growth, and division affect population-level quantities
such as population-averaged cell size.  The kinetic theory PDE can be
marginalized in different ways and reduce, in different limits, to
master-like equations or structured population-like PDEs, thus
unifying deterministic ``moment'' equations (the structured population
PDEs) with Markovian birth-death-like models. Stochastic fluctuations
in parameters such as the cellular growth rate have also been included
\cite{AMIR_REVIEW}, but integrating fluctuations of internal variables
with random birth-death events (demographic stochasticity) is
challenging due to the combinatorial complexity and unwieldiness of
the relevant equations.

Besides simple individual-cell dynamical variables such as cell age or
cell size, gene (mRNA) or protein expression levels are also measured
cellular attributes that are important in cell biology, particularly
during development. Since there are many different species of mRNA or
proteins, the expression pattern is a vector of fluctuating variables.

Although modern computational and statistical techniques can be used
to quantitatively infer single cellular mRNA \cite{la2018rna} or
protein \cite{QIU2022,gorin2020protein} levels from experimental data,
mathematical models of how expression levels or cell states evolve is
often couched in terms of transport along Waddington or fitness
landscapes
\cite{bhattacharya2011deterministic,wang2011quantifying}. The value of
the landscape may represent an ``energy'' function that is shaped by
different genes, or a proliferation rate that is a function different
gene expression rates. However, how populations of cells are
represented in such high-dimensional ``landscapes'' is unclear.
Moreover, since cellular division rates and death rates typically
depend in depend on internal stochastic cell variables such as gene
expression levels
\cite{peng2022regulated,marzluff2002histone,heintz1983regulation},
it is important to model how fluctuating-gene-expression-dependent 
birth or death rates feature in the evolution of 
a population along an appropriate landscape. 

Kinetic models have the capability of precisely describing both the
stochastic dynamics of individual cell states and the stochastic
birth-death processes associated with an evolving population. Not only
is the coupling between individual cell states and the evolution of
the population explicit in a kinetic equation, but potential functions
governing intracell state dynamics and proliferation (defining a
fitness function) arise naturally in the kinetic framework.

Previously derived kinetic models such as the timer-sizer model for
cell populations distributed across size \cite{Xia2020,Xia_2021}
incorporate stochastic differential equations (SDEs) to track the
dynamics individual internal cell states such as size or mRNA/protein
levels.  Marginalization of the kinetic equations results in equations
for the correlation functions that explicitly show how individual cell
states are linked to key macroscopic quantities of the overall
population. However, these kinetic theories could not track lineages
or generational subpopulations of cells nor did they incorporate cell
death or cell division that may also depend on other stochastic
variables associated with the cell.

In this paper, we formally develop a complete kinetic model that
tracks continuous-valued, stochastically evolving variables
(\textit{e.g.}, gene expression, cellular size, mRNA level, protein
level, etc.)  and the discrete generation number of each cell.  The
mathematical framework we use for delineating cell of differernt
attribute values across different generations shares a related
structure to one recently used to describe ages across different cell
stages \cite{YATES2022}. In our problem, noise in gene expression is
described by a continuous-time stochastic process while noise in
division events is described by a Markov jump process.  Our model
couples these stochastic processes through an SDE-jump-process hybrid
model in which the division and death rates explicitly depend on
fluctuating gene expression levels
\cite{buccitelli2020mrnas,xia2019spatial}. All of these quantities are
tracked along different generations. The mathematical framework we use
for delineating cell of differernt attribute values across different
generations shares a related structure to one recently used to
describe ages across different cell stages \cite{YATES2022}.
 



In the next section, we define the kinetic model and show how
potentials that govern the intracellular dynamics and the population
fitness can be motivated. Since the development of our
generation-dependent kinetic equations requires intensive book-keeping
and associated notation to resolve the time-dependent attributes of
each member of the entire population, many of the steps are detailed
in extensive mathematical Appendices.  However, eventually, in
Section~\ref{meanfield} we marginalize our high-dimensional kinetic
PDE to derive a number of more meaningful ``reduced'' equations that
describe the evolution of key quantities of biological interest. These
new results are summarized and listed in the Summary and Conclusions.
We also carry out a numerical experiment on a simple example to show
how cellular gene expression levels evolve over generations and how
the macroscopic cellular density (with respect to gene expression
level), when interrupted by cellular division, can be prevented from
returning to the equilibrium distribution. In the Conclusions, we
discuss potential applications and extensions.

\section{Kinetic equation formulation}
\label{sec:derivation}

For simplicity, we first assume the internal state of each cell is
characterized by a one-dimensional scalar quantity $X\in\mathbb{R}$.
This continuous stochastic variable may represent, for example, the
expression level of a single mRNA transcript or protein abundance (or
log-abundance). Besides this continuous variable, associated with each
cell is the discrete generation $i\in\mathbb{N}^+$ to which it belongs
(assuming it is part of a lineage derived from an ancestor).
%
%
%
We model the evolution of $X_{i, j}$ (the internal state of the
$j^{\text{th}}$ cell in the $i^{\text{th}}$ generation) using an SDE
of the standard form \cite{crispin2009stochastic,coomer2020shaping}
\begin{equation}
\dd X_{i, j}(t) = g_{i, j}(X_{i, j}, t)\dd{t} + \sigma_{i, j}(X_{i, j}, t)\dd W_{i, j},
\label{Xevolve}
\end{equation}
%
%
where $g_{i, j}(X_{i, j}, t)\dd{t}$ is the deterministic convection
that depends on both $X_{i, j}$ and the generation $i$, and $\dd W_{i,
  j}$ are increments of independent Wiener processes for each $i,
j$. Thus, the term $\sigma_{i, j}(X_{i, j}, t)\dd W_{i, j}$ represents
the ``intrinsic'' fluctuation in the evolution of $X_{i,
  j}(t)$. Often, one can assume that the convection arises from
gradients of a potential ``energy function'' $\Phi$: $g_{i, j}(X_{i,
  j}, t)\coloneqq -\nabla \Phi(x,t)|_{x=X_{i, j}}$
\cite{wang2011quantifying}. Although a gradient of $\Phi(x,t)$ may
conveniently describes a time-dependent force that changes gene
expression, nonconservative driving with metabolically driven fluxes,
which cannot be described by a potential, is also to be expected
\cite{wang2015landscape}.

    
We assume that both $g_{i}$ and $\sigma_{i}$ are Lipschitz continuous
so the solution $X_{i, j}(t)$ of Eq.~\eqref{Xevolve} exists and is
almost surely unique given any initial condition $X_{i, j}(0)$. The
evolution of $X_{i, j}$ is interrupted by the cell division; an
$i^{\text{th}}$ generation cell with internal state $X_{i, j}$ divides
in time $\dd t$ with total probability $\beta_{i}(X_{i, j})\dd t$.
This Markovian birth rate can be further stratified by internal state
of the two resulting daughter cells immediately after their birth.  We
denote the differential birth rate density of producing one daughter
with internal state $X_1$ and the other with state $X_2$ as
$\tilde{\beta}_{i, j}(X_{i, j}, X_1, X_2)$. Integrating over all
possible daughter cell states $X_{1}, X_{2}$ defines the total division
rate:
\begin{equation}
  \int \tilde{\beta}_{i, j}(X_{i, j}, X_1, X_2)\dd{X}_1\dd{X}_2 = \beta_{i, j}(X_{i, j}).
\end{equation}
A form for $\tilde{\beta}$ might be
\begin{equation}
\tilde{\beta}_{i, j}(X_{i, j}, X_1, X_2) \propto e^{-\phi(X_1, X_2|X_{i, j})},
\end{equation}
which defines a ``free energy'' function $\phi(X_1, X_2|X_{i, j})$ for
the rate of a mother cell with attribute value $X$ to divide into
daughters cell with attribute values $X_1$ and $X_2$. If the states of
the daughter cells tend towards being similar in value to that of
their mother cell, then $\phi(X_1, X_2|X_{i, j})$ would exhibit a
minimum at $X_1, X_2\approx X$. Although $\Phi$ and $\phi$ might be
loosely described in terms of Waddington and fitness landscapes, our
unifying kinetic framework allows them to be unambiguously described
in terms of the intracellular advection $g_{i,j}(X_{i,j}, t)$ and
proliferation function $\tilde{\beta}$, respectively.


Since the derivation of our kinetic theory requires the use of a
number of variables and indices, we define some simplifying
notation. Specifically, each of the $n_{i}$ elements of the bold
vector $\bm{X}_i$ represents the expression level $X_{i,j}$ of the
$j^{\rm th},\, 1\leq j \leq n_{i}$ cell in the $i^{\rm th}$-generation
subpopulation. These vectors $\bm{X}_i$ for the subpopulations across
generations $1\leq i\leq k$ can be collected as a matrix defined as
$\bm{X}_{\bm{n}}\coloneqq(\bm{X}_1, \ldots, \bm{X}_k)$, where
$\bm{n}\coloneqq(n_1, ..., n_k)$ is a vector representing the total
number of cells in each generation $1\leq i\leq k$. Each value $n_{i}$
evolves stochastically defined by random birth and death events. Below
is a table of the various definitions and overall notation used
throughout this paper.
\begin{table*}[t]
\centering
\renewcommand*{\arraystretch}{1.1}
\begin{tabular}{| >{\centering\arraybackslash} m{9em}| 
>{\arraybackslash} m{40em}|}\hline \textbf{Symbol} &
  \textbf{Definition and explanation} \\[1pt]
  \hline\hline
 \,\,\, $\bm{n}(t)$\,\, & $\bm{n}(t)\coloneqq(n_1(t), ...,
  n_{k(t)}(t))$: time-dependent vector of random numbers of cells in the $i^{\text{th}}$
  generation, $i=1,...,k$ \\[2pt] \hline
  \,\,\, $\bm{n}$\,\, & $\bm{n}\coloneqq(n_1, ..., n_k)$: vector of
  integer values $n_{i}$ of the number of cells in generation $i=1,...,k$ \\[2pt]
  \hline
  \,\,\,  $\bm{X}(t)_{\bm{n}(t)}$\,\, &
  $\bm{X}(t)_{\bm{n}(t)}\coloneqq(\bm{X}_1(t), ...,
  \bm{X}_{k(t)}(t))$, $\bm{X}_i(t)\coloneqq(X_{i, 1}(t),...,X_{i,
    n_i(t)}(t))$: time-dependent random variable describing the state of each cell,
  \textit{e.g.,} gene expression level $X_{i, n_{i}(t)}$ of the $n_{i}^{\rm th}$ cell in the $i^{\rm th}$ generation
  \\[2pt] \hline
  \,\,\, $\bm{X}_{\bm{n}}$\,\, & $\bm{X}_{\bm{n}}\coloneqq(\bm{X}_1,
  ..., \bm{X}_k)$, $\bm{X}_i\coloneqq(X_{i, 1},...,X_{i, n_i})$:
  values of $\bm{X}(t)_{\bm{n}(t)}$ \\[2pt] \hline
  %
  %
  \,\,\, $\vec{X}_{n}$\,\, & $\vec{X}_{n}
    \coloneqq(X_1, ..., X_n)$,
  the vector of state values for any collection of $n$ cells \\[2pt]
  \hline
  \,\,\, $g_{i, j}(X_{i, j}, t)$\,\, & deterministic growth rate
  of the $j^{\text{th}}$ cell in the $i^{\text{th}}$ generation
  \\[2pt] \hline
  \,\,\, $\sigma_{i, j}(X_{i, j}, t)$\,\, & noise in
  the growth of the $j^{\text{th}}$ cell in the $i^{\text{th}}$
  generation \\[2pt] \hline
  \,\,\, $\beta_{i, j}(X_{i, j})$\,\, & division rate of the
  $j^{\text{th}}$ cell in the $i^{\text{th}}$ generation \\[2pt]
  \hline
  \,\,\, $\mu_{i, j}(X_{i, j})$\,\, & death rate of the
  $j^{\text{th}}$ cell in the $i^{\text{th}}$ generation \\[2pt]
  \hline
  \,\,\, $\tilde{\beta}_{i, j}(X_{i, j}, X_1, X_2)$\,\, &
  differential division rate of the $j^{\text{th}}$ cell in the
  $i^{\text{th}}$ generation into two cells in the
  $(i+1)^{\text{th}}$ generation with states $X_1, X_2$ \\[2pt] \hline
  \,\,\, $\bm{X}_{\bm{n}_{\text{b}, -i}}^{{-j}}$\,\, &
  states of the cell population right after the
  $j^{\text{th}}$ cell in the $i^{\text{th}}$ generation
  divides. $\bm{X}_{\bm{n}_{\text{b},
      -i}}^{{-j}}$ differs from $\bm{X}_{\bm{n}}$ in that the state
  variables for the cells in the $(i-1)^{\text{th}}$ generation is
  $(X_{i-1, 1},...,X_{i-1, j-1},X_{i-1, j+1},...X_{i-1, n_i})$ and the
  state variables for the cells in the $i^{\text{th}}$ generation are
  $(X_{i, 1},...,X_{i, n_i}, X_1, X_2)$ \\[2pt] \hline
  \,\,\, $\bm{X}^{-j}_{\bm{n}_{\text{d}, -i}}$\,\, &
  states of the cell population right after the $j^{\text{th}}$ cell
  in the $i^{\text{th}}$ generation dies. $\bm{X}^{-j}_{\bm{n}_{\text{d}, -i}}$
  differs from $\bm{X}_{\bm{n}}$ in that the state variables for the
  cells in the $(i-1)^{\text{th}}$ generation are $(X_{i-1,
    1},...,X_{i-1, j-1},X_{i-1, j+1},...X_{i-1, n_i})$ \\[2pt] \hline
  \,\,\, $\bm{X}_{\bm{n}_{\text{b}, i-1}}^j$\,\, & pre-division
  cellular population: it differs from $\bm{X}_{\bm{n}}$ in that the
  state variables for the cells in the $(i-1)^{\text{th}}$ generation
  is $(X_{i-1, 1},...,X_{i-1, j-1}, Y, X_{i-1, j},...)$ and the state
  variables for the cells in the $i^{\text{th}}$ generation are
  $(X_{i, 1},...,X_{i, n_i-2})$ (an additional cell with $Y$ in the
  $(i-1)^{\text{th}}$ generation divides and gives birth to two new
  daughter cells $X_{i, n_i-1}, X_{i, n_i}$ in the $i^{\text{th}}$
  generation)\\[2pt] \hline
  \,\,\, $\bm{X}_{\bm{n}_{\text{d}, i}}^j$\,\, & pre-death
  cell population states. This differs from $\bm{X}_{\bm{n}}$ in that the
  state variables for the cells in the $i^{\text{th}}$ generation are
  $(X_{i, 1},...,X_{i, j-1}, Y, X_{i, j},...)$ (an additional cell in
  the $i^{\text{th}}$ generation with $Y$ dies) \\[2pt] \hline
  \,\,\, \added{$\bm{X}_{\bm{n}_{\text{b}, i}}^{j_1, j_2}$} & \added{pre-division
  state which differs from $\bm{X}_{\bm{n}}$ in that the state vector
  associated with the $i^{\text{th}}$ generation is $(Y, X_{i,
    1},...X_{i, n_{i}})$ and the state of the $(i+1)^{\text{th}}$
  generation does not contain components $X_{i+1, j_1}$ and $X_{i+1,
    j_2}$}  \\[2pt] \hline
\end{tabular}
\vspace{1mm}
\caption{\small\textbf{Overview of variables.} A list of the main
  variables and parameters used. \added{The specific labels and
    definitions of state vectors given provide the proper bookkeeping
    of all possible initial and final states upon birth and death.}}
\label{tab:model_variables}
\end{table*}

Next, define $p_{\bm{n}}(\bm{X}_{\bm{n}}, t|\bm{X}(0)_{\bm{n}(0)}, 0)$ as
the probability density function that the population has $\bm{n}$
cells with internal states $\bm{X}_{\bm{n}}$
given the initial condition that the system has $\bm{n}(0)$
%
%
cells with internal state values $\bm{X}(0)_{\bm{n}(0)}$ at $t=0$. For
notational simplicity, we name the cell state random variables (at
time $t$) $X_{i, j}(t), \bm{X}_{i}(t)$, and $\bm{X}(t)_{\bm{n}(t)}$, and
denote their values by and $X_{i, j}, \bm{X}_i$, and
$\bm{X}_{\bm{n}}$, respectively.  The probability density
$p_{\bm{n}}(\bm{X}_{\bm{n}}, t \vert \bm{X}(0)_{\bm{n}(0)}, 0)$ can be
defined as the expectation over trajectories from
($\bm{X}(0)_{\bm{n}(0)}, 0$) to ($\bm{X}_{\bm{n}}, t$):


\begin{widetext}
  
\begin{equation}
  p_{\bm{n}}\big(\bm{X}_{\bm{n}}, t|\bm{X}(0)_{\bm{n}(0)}, 0\big) =
  \begin{cases}
    \begin{aligned}
      & \mathbb{E}\Big[\delta\big(\bm{X}(t)_{\bm{n}(t)}
        - \bm{X}_{\bm{n}}\big) S\big(t;\bm{X}(t)_{\bm{n}(t)}\big)
        \Big|\bm{X}(0)_{\bm{n}(0)},0; \bm{n}(0<s<t)=\bm{n}(0)\Big], \quad\,\,
      \bm{n} = \bm{n}(0)  \\[-6pt]
      & \hspace{2mm} + \!
      \int_0^t \!\mathbb{E}\Big[\tilde{J}\big(t, \tau;\bm{X}(t)_{\bm{n}(t)},\bm{n}(0)\big)
        S\big(\tau;\bm{X}(\tau)_{\bm{n}(\tau)}\big)
        \Big|\bm{X}(0)_{\bm{n}(0)}, 0; \bm{n}(0<s<\tau)=\bm{n}(0)\Big] \dd \tau,
    \\[12pt]
   & \displaystyle\mathbb{E}\Big[\int_0^t
     \tilde{J}(t,\tau; \bm{X}(0)_{\bm{n}(0)}, \bm{n}(0))
S\big(\tau; \bm{X}(\tau)_{\bm{n}(\tau)}\big)\, \dd\tau 
\Big|\bm{X}(0)_{\bm{n}(0)}, 0\Big], \qquad \qquad\quad\,\, \bm{n} \neq \bm{n}(0)
    \end{aligned}
  \end{cases}
\label{pdef}
\end{equation}
where

\begin{equation}
\begin{aligned}
  %
  S\big(t; \bm{X}(t)_{\bm{n}(t)}\big)\equiv & \exp
  \Big[-\!\int_{0}^{t}\sum_{i=1}^{k(0)}\sum_{j=1}^{n_{i}(0)}
\Big(\beta\big(X_{i, j}(s)\big)+\mu\big(X_{i, j}(s)\big)\Big)\dd s\Big] \\
\tilde{J}\big(t, \tau;\bm{X}_{\bm{n}},\bm{n}(0)\big) \equiv &
\sum_{i=1}^{k(0)}\sum_{j=1}^{n_{i}(0)}\Big[\tilde{\beta}_{i, j}
  \big(X_{i, j}(\tau), X_1(\tau), X_2(\tau)\big)
p_{\bm{n}}(\bm{X}_{\bm{n}}, t-\tau|\bm{X}(\tau)_{\bm{n}(0)_{\text{b}, -i}}^{-j}, 0) \\[-3pt]
\: & \qquad \hspace{3cm}+ \mu_{i, j}\big(X_{i, j}(\tau)\big)p_{\bm{n}}(\bm{X}_{\bm{n}}, t-\tau|
\bm{X}(\tau)_{\bm{n}(0)_{ \text{d}, -i}}^{-j}, 0)\Big].
\end{aligned}
\label{Wdef}
\end{equation}

Definitions of $\bm{X}^{-j}_{\bm{n}(0)_{ \text{b}, -i}}(s)$ and
$\bm{X}^{-j}_{\bm{n}(0)_{\text{d}, -i}}$ are given in
Table~\ref{tab:model_variables}.
%
%
The term $S\big(t; \bm{X}(t)_{\bm{n}(t)}\big)$ represents the survival
probability up to time $t$ while $\tilde{J}(t, \tau;
\bm{X}(t)_{\bm{n}(t)}, \bm{n}(0))$ describes the probability flux from
a given state $\bm{X}(\tau)_{\bm{n}(\tau)}$ to the current state
$\bm{X}(t)_{\bm{n}(t)}$ due to division or death at time $\tau$.  The
first form on the RHS of Eq.~\eqref{pdef} is the probability that no
division or death happens in the system during time $[0, t]$ and the
final internal states of the cell population are
$\bm{X}(t)_{\bm{n}(t)}$ while the second form in Eq.~\eqref{pdef}
denotes the probability that at least one division or death happened
within $[0, t]$ to arrive at the final internal state
$\bm{X}(t)_{\bm{n}(t)}$.

We shall show that under certain conditions,
$p_{\bm{n}}(\bm{X}_{\bm{n}}, t|\bm{X}(0)_{\bm{n}(0)}, 0)$ satisfies
the partial differential equation

\begin{equation}
\begin{aligned}
  \f{\p p_{\bm{n}}}{\p t} + \sum_{i=1}^k\sum_{j=1}^{n_i}\f{\p(g_{i, j}
    p_{\bm{n}})}{\p X_{i, j}} = & 
\mfrac{1}{2} \sum_{i=1}^k\sum_{j=1}^{n_i}\f{\p^2 (\sigma_{i, j}p_{\bm{n}})}{(\p X_{i, j})^2}  - 
\sum_{i=1}^k\sum_{j=1}^{n_i}\big(\beta_{i, j}(X_{i, j})+\mu_{i, j}(X_{i, j})\big) p_{\bm{n}} \\ 
\: & + \sum_{i=2}^k\sum_{j=1}^{n_{i-1}+1}\!\int\! \tilde{\beta}(Y, X_{i, n_i-1}, X_{i, n_i})
p_{\bm{n}_{\text{b}, i-1}}(\bm{X}^j_{\bm{n}_{\text{b}, i-1}}, t|\bm{X}(0)_{\bm{n}(0)}, 0)\,\dd{Y} \\
\: & + \sum_{i=1}^{\infty}\sum_{j=1}^{n_i+1}\!\int\! 
\mu(Y)p_{\bm{n}_{\text{d}, i}}(\bm{X}^j_{\bm{n}_{\text{d}, i}}, t|\bm{X}(0)_{\bm{n}(0)}, 0)\, \dd{Y}.
\end{aligned}
\label{ppde2}
\end{equation}
In Eq.~\eqref{ppde2}, the pre-division cell population
$\bm{X}_{\bm{n}_{\text{b}, i-1}}^{{j}}$ and the pre-death cell
population $\bm{X}_{\bm{n}_{\text{d}, i}}^{{ j}}$ are explicitly
defined in Table~\ref{tab:model_variables}.  The mathematical steps
and necessary conditions needed to show that
$p_{\bm{n}}(\bm{X}_{\bm{n}}, t|\bm{X}(0)_{\bm{n}(0)}, 0)$ defined in
Eq.~\eqref{pdef} satisfies Eq.~\eqref{ppde2} is given in
Appendix~\ref{derivationsde}. We impose the normalization condition
$\sum_{{\bm{n}}}\int p_{\bm{n}}(\bm{X}_{\bm{n}},
t|\bm{X}(0)_{\bm{n}(0)}, 0) \dd{\bm{X}_{\bm{n}}}= 1$ for every
$\bm{X}(0)_{\bm{n}(0)}$ and average over an initial distribution of
$\bm{X}(0)_{\bm{n}(0)}(0)$ (denoted by
$q_{\bm{n}(0)}(\bm{X}(0)_{\bm{n}(0)}, 0)$) to define an unconditional
probability density

\begin{equation}
\begin{aligned}
  p_{\bm{n}}(\bm{X}_{\bm{n}}, t)\coloneqq \sum_{\bm{n}(0)}\int_{\bm{X}_{\bm{n}(0)}}\!\!
  p_{\bm{n}}(\bm{X}_{\bm{n}}, t|\bm{X}(0)_{\bm{n}(0)}, 0)
  q_{\bm{n}(0)}(\bm{X}(0)_{\bm{n}(0)}, 0)\, \dd \bm{X}(0)_{\bm{n}(0)}
\end{aligned}
\label{psdef}
\end{equation}
that also satisfies Eq.~\eqref{ppde2}.

Next, we define the symmetric probability density distribution

\begin{equation}
  \rho_{\bm{n}}(\bm{X}_{\bm{n}}, t) \coloneqq \prod_{i=1}^k \frac{1}{n_i!}
  \sum_{\pi} p_{\bm{n}}(\pi(\bm{X}_{\bm{n}}), t)
\label{rhodef}
\end{equation}
where $p_{\bm{n}}$ is defined in Eq.~\eqref{psdef} and
$\pi(\bm{X}_{\bm{n}})$ is a permutation operator that reorders the
sequence of the state variables $X_{i, j}$ of cells within each
generation, for all generations. Thus, the summation is taken over all
such grouped permutations ($\prod_{i=1}^k {n_i!}$ permutations in
total). In the special case

\begin{equation}
  g_{i, j}=g_i, \, \sigma_{i, j}=\sigma_i, \,
  \beta_{i, j}=\beta_i,\,  \mu_{i, j}=\mu_i, \, \tilde{\beta}_{i, j}=\tilde{\beta}_i,
\end{equation}

\textit{i.e.}, when the rate parameters depend at most on the
generation of a cell, $\rho_{\bm{n}}(\bm{X}_{\bm{n}}, t)$ defined in
Eq.~\eqref{rhodef} obeys

\begin{equation}
\begin{aligned}
  \f{\p \rho_{\bm{n}}}{\p t} + \sum_{i=1}^k\sum_{j=1}^{n_i}
  \f{\p(g_{i}\rho_{\bm{n}})}{\p X_{i, j}} = & 
\mfrac{1}{2} \sum_{i=1}^k\sum_{j=1}^{n_i} \f{\p^2 (\sigma_{i, j}\rho_{\bm{n}})}{(\p X_{i, j})^2} 
-\sum_{i=1}^k\sum_{j=1}^{n_i}\big(\beta_{i, j}(X_{i, j})+\mu_{i, j}(X_{i, j})\big) \rho_{\bm{n}} \\ 
\: & +\sum_{i=1}^{k-1}\frac{n_{i}+1}{n_{i+1}(n_{i+1}-1)}\!\sum_{1\leq j_1\neq j_2\leq n_{i+1}}
\!\!\!\medint\int \tilde{\beta}_i(Y, X_{i+1, j_1}, X_{i+1, j_2})
\rho_{\bm{n}_{\text{b}, i}}(\bm{X}_{\bm{n}_{\text{b}, i}}^{{j_1, j_2}}, t)\,\dd{Y} \\
\: & + \sum_{i=1}^{\infty}\sum_{j=1}^{n_i+1}\!\int\!\mu_i(Y)
\rho_{\bm{n}_{\text{d}, i}}(\bm{X}_{\bm{n}_{\text{d}, i}}^{{j}}, t)\,\dd{Y},
\end{aligned}
\label{symmeqn}
\end{equation}
where $\bm{X}_{\bm{n}_{\text{b}, i}}^{j_1, j_2}$ differs from
$\bm{X}_{\bm{n}}$ in that the state vector associated with cells in
the $i^{\text{th}}$ generation are $(Y, X_{i, 1},...X_{i, n_{i}})$ and
the state vector for cells in the $(i+1)^{\text{th}}$ generation does
not have the components $X_{i+1, j_1}$ and $X_{i+1, j_2}$.

Finally, in many systems, the state variable is a multi-dimensional
vector instead of a scalar, \textit{i.e.}, $X_{i, j}\coloneqq(X_{i, j,
  1},..., X_{i, j, d})\in\mathbb{R}^d$ may also represent $d$
different gene or protein expression levels in the $j^{\text{th}}$
cell in the $i^{\text{th}}$ generation.  This vector may represent,
for example, $d$ different gene or protein expression levels.  We
assume that the evolution of $X_{i, j}$ (each element now implicitly a
vector of attributes) follows the Brownian SDE

\begin{equation}
\dd X_{i, j} = \bm{g}_{i, j}(X_{i, j}, t)\dd{t} 
+ \bm{\Sigma}_{i, j}(X_{i, j}, t)\dd \bm{W}_{i, j}
\label{SDE_d}
\end{equation}
where $\bm{W}_{i, j}$ is a $d_0$-dimensional vector of independent
Wiener processes ($d_{0} \leq d$) for each $i, j$ and the coefficients
$\bm{g}_{i, j}(X_{i, j}, t)\coloneqq(g_{i, j, 1}(X_{i, j},
t),...,g_{i, j, d}(X_{i, j},
t)):\mathbb{R}^{d}\times\mathbb{R}^+\rightarrow\mathbb{R}^d,
\big(\bm{\Sigma}_{i, j}\big)_{mn} \coloneqq \big(\sigma_{i,
  j}(X_{i,j},t)\big)_{mn}:
\mathbb{R}^{h}\times\mathbb{R}^+\rightarrow\mathbb{R}^{d\times d_0},
m=1,...,d, n=1,...,d_0$ \added{are all smooth, uniform Lipschitz continuous,
and uniform bounded. We can also define the symmetric probability
density distribution $\rho_{\bm{n}}(\bm{X}_{\bm{n}}, t)$ as in
Eqs.~\eqref{rhodef} and after applying the multi-dimensional forward
Feynman-Kac equation case in \cite{lecavil2015probabilistic} we can
show that the differential equation satisfied by such $\rho_{\bm{n}}$
is}\\

\begin{equation}
\begin{aligned}
  \f{\p \rho_{\bm{n}}}{\p t} + \sum_{i=1}^k\sum_{j=1}^{n_i}\sum_{\ell=1}^d
  \f{\p(g_{i,j, \ell}\rho_{\bm{n}})}{\p X_{i, j, \ell}} = & \mfrac{1}{2} \sum_{i=1}^k
  \sum_{j=1}^{n_i}\sum_{\ell_1, \ell_2=1}^{d}\!\!
  \f{\p^2 (\sum_{h=1}^{d_{0}}(\Sigma_{i,j})_{\ell_1, h}(\Sigma_{i,j})_{\ell_2, h}
   \, \rho_{\bm{n}})}{(\p X_{i, j, \ell_1}\p X_{i, j, \ell_2})}\\
  \: & \,\, -\sum_{i=1}^k\sum_{j=1}^{n_i}
  \big(\beta_{i, j}(X_{i, j})+\mu_{i, j}(X_{i, j})\big)\rho_{\bm{n}} \\
  \: & \,\,  \sum_{i=1}^{k-1}\frac{n_{i}+1}{n_{i+1}(n_{i+1}-1)}\!
  \sum_{1\leq j_1\neq j_2\leq n_{i+1}}\!\!\!\!\!\medint\int
  \tilde{\beta}(Y, X_{i+1, j_1},X_{i+1, j_2})
  \rho_{\bm{n}_{\text{b}, i}}(\bm{X}_{\bm{n}_{\text{b}, i}}^{j_1, j_2}, t)\,\dd{Y} \\
  \: & \,\, + \sum_{i=1}^{\infty}\sum_{j=1}^{n_i+1}\medop\int\mu_{i, j}(Y)
  \rho_{\bm{n}_{\text{d}, i}}(\bm{X}_{\bm{n}_{\text{d}, i}}^{j}, t)\, \dd{Y}
  \label{symmeqn_vector}
\end{aligned}
\end{equation}
if the coefficients $\beta_{i, j}=\beta_i, \mu_{i, j}=\mu_i,
\tilde{\beta}_{i, j}=\tilde{\beta}_i$ are homogeneous for cells in the
same generation.

In Appendix \ref{appendixd}, we also derive kinetic equations for the
population density associated with cells that are also labeled by
their age. The derivation assumes the budding model of birth where on
daughter cell's age is set to zero immediately after birth
\cite{chou2016hierarchical_PRE,chou2016hierarchical}. 

\end{widetext}

\section{Mass-action differential equations}
\label{meanfield}
Henceforth, we will consider the ``simpler'' single-gene
model. Extension to $d$-dimensional attributes can be implemented
following the structure in Eqs.~\eqref{SDE_d} and
\eqref{symmeqn_vector}.

Through marginalization of the kinetic equation~\eqref{symmeqn} we can
derive he differential equations that describe the evolution of
certain ``macroscopic'' quantities such as the expected
total-population levels of $X$. In this section, we derive governing
equations for examples of macroscopic quantities by marginalizing
Eq.~\eqref{symmeqn}, which are then solved numerically to show how
quantities such as cellular gene expression levels can evolve over
generations.

\subsection{Evolution of the population density}
First, we can track the marginal cell distributions of certain cells in
specified generations by defining the macroscopic quantity

\begin{equation}
  u_{\bm{n}}(\bm{X}_{\bm{n}}, t) \coloneqq
  \!\sum_{\bm{m}\geq \bm{n}}\prod_{\ell=1}^{\infty}(\bm{m}_{\ell})_{\bm{n}_{\ell}}
  \int_{\bm{X}_{\bm{m}\backslash \bm{n}}}\!\!\!\rho_{\bm{m}}(\bm{X}_{\bm{m}}, t)
  \dd\bm{X}_{\bm{m}\backslash \bm{n}},
\label{u_definition}
\end{equation}
where $\bm{m}\geq \bm{n}$ means that for each component in
$\bm{m}\coloneqq(m_1,...,m_{\ell}), m_{\ell} \geq n_{\ell}$ and
$(m_{\ell})_{n_{\ell}}\coloneqq
m_{\ell}(m_{\ell}-1)...(m_{\ell}-n_{\ell}+1)$ is the falling
factorial.  The integration is taken over the remaining variables
$\bm{X}_{\bm{m}}$, but excludes the variables of interest
$\bm{X}_{\bm{n}}$ which are retained. We find that
$u_{\bm{n}}(\bm{X}_{\bm{n}}, t)$ satisfies the differential equation

\begin{equation}
\begin{aligned}
  \f{\p u_{\bm{n}}}{\p t} + \sum_{i=1}^k\sum_{j=1}^{n_i}\f{\p(g_{i}u_{\bm{n}})}{\p X_{i, j}} = &
  \mfrac{1}{2} \sum_{i=1}^k\sum_{j=1}^{n_i} \f{\p^2 (\sigma_{i}u_{\bm{n}})}{(\p X_{i, j})^2} - 
  \sum_{i=1}^k\sum_{j=1}^{n_i}\big(\beta_i(X_{i, j})+\mu_i(X_{i, j})\big) u_{\bm{n}} \\
  \: & \quad  + \sum_{i=1}^{k-1}\sum_{j_1\neq j_2}\!\medint\int\tilde{\beta}_{i, j}(Y, X_{i+1, j_1}, X_{i+1, j_2})
   u_{\bm{n}_{\text{b}, i}}(\bm{X}_{\bm{n}_{\text{b}, i}}^{{ j_1, j_2}}, t)\, \dd{Y} \\
  \: & \quad + \sum_{i=1}^{k-1}\sum_{j=1}^{n_{i+1}}
  \medint\int\Big(\tilde{\beta}_{i, j}(Y, X_{i+1, j}, Z)
  + \tilde{\beta}_{i}(Y, Z, X_{i+1, j})\Big)
  u_{\bm{n}_{\text{b}, i}}(\bm{X}_{\bm{n}_{\text{b}, i}}^{{ j}}, t)\,\dd Y\dd Z.
\end{aligned}
\label{u_structured}
\end{equation}


From Eq.~\eqref{u_structured}, the set of macroscopic quantities
$\{u_{\bm{n}}\}$ satisfies ``sequential'' closed-form equations in
that the PDE satisfied by $u_{\bm{n}}$ depends only on
$u_{\bm{n}_{\text{b}, i}}(\bm{X}_{\bm{n}_{\text{b}, i}}^{{ j_1, j_2}},
t)$ and $u_{\bm{n}_{\text{b}, i}}(\bm{X}_{\bm{n}_{\text{b}, i}}^{{
    j}}, t)$.  In the specific case $\bm{n}_i\coloneqq(0, .., 0,
1)\in\mathbb{R}^i$, $u_{\bm{n}_i}(\bm{X}_{\bm{n}_i}, t)$ tracks the
$i^{\text{th}}$-generation cell population density in the structured,
one-dimensional variable $X_{i, 1}$. The quantity
$\{u_{\bm{n}_i}(\bm{X}_{\bm{n}_i}, t)\}_{i=1}^{\infty}$ indicates how
the cellular population density evolves across generations through
division and differentiation.

Consider the specific example studied in \cite{coomer2022noise} where
the coefficients in Eq.~\eqref{u_structured} take the form
\begin{equation}
g_{i, j}(X_{i, j}, t) = -X_{i, j}, \quad \sigma^2_{i, j}(X_{i, j}, t) = \exp(-X_{i, j}^2).
\end{equation}
In this case, if the cells do not divide or die (\textit{i.e.}, the
entire population stays in the stay first generation), and their
attributes converge to an equilibrium distribution
\begin{equation}
  u_{\bm{n}_1}(X_{1,1}=x, t\to \infty)
  = \frac{\exp\big[2x^2-\tfrac{1}{2}e^{2x^{2}}\big]}{Z},
  \label{u_inf}
\end{equation}
where $Z = \int_{-\infty}^{\infty}\exp[2x^2-\tfrac{1}{2}e^{2x^{2}}]\dd x$ is
the normalization constant.

\begin{figure*}[htbp]
\centering
\includegraphics[width=\textwidth]{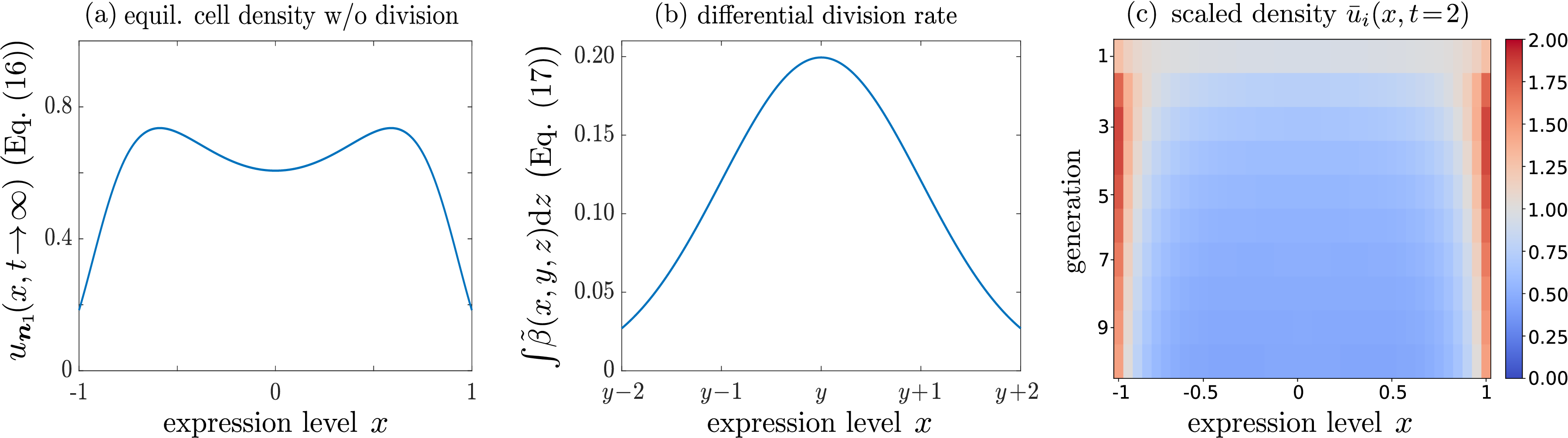}
\caption{\small (a) The equilibrium cellular density without division
  (Eq.~\eqref{u_inf}).  (b) A differential birth rate
  $\int\tilde{\beta}_{i,j}(X_{i, j}, Y, Z)\dd Z$ using the form given
  in Eq.~\eqref{betaij}. (c) Using the differential birth rate in (b)
  and Eq.~\eqref{u_inf} for normalization, we plot the associated
  cellular density $\bar{u}_{i}(x,t=2)$ (Eq.~\eqref{ubar}) across
  different generations. The differentiation process prevents the
  population from reaching an equilibrium ($i\geq 2$) even when the
  death rate and division rate are $x$-independent.  However, as time
  increases for a certain generation (such as $i=1$) in which no cell
  has entered, the structured population in that generation gradually
  returns to equilibrium.}
\label{oscillatory}
\end{figure*}

To include birth and death, we choose birth and death rates of the form
\begin{equation}
\beta_{i, j}=\mfrac{1}{2},\quad \mu_{i, j} = \frac{i-1}{2i}, \quad
\int \tilde{\beta}_{i, j}(X_{i, j}, Z, Y)\dd Z =
\int\tilde{\beta}_{i, j}(X_{i, j},Y,  Z)\dd Z
\equiv \frac{\beta_{i}}{\sqrt{2\pi}}e^{-\frac{(Y-X_{i, j})^2}{2}}
\label{betaij}
\end{equation}
and set the initial condition to be $u_{\bm{n}_i}(\{X\}_{\bm{n}_i},
t)=\big(\tfrac{1}{100}\big)\delta_{i, 1}\, \mathbbm{1}_{-2.5\leq X_{1,
    1}\leq 2.5}$, where $\delta_{i, j}=1$ if $i=j$ and $\delta_{i,
  j}=0$ otherwise is the Kroenecker $\delta$-function and
$\mathbbm{1}$ is the indicator function). Using these parameters and
initial condition, we plot the scaled (using Eq.~\eqref{u_inf})
generation-dependent cellular density

\begin{equation}
\bar{u}_{i}(x,t)\equiv
\left(\frac{1}{u_{\bm{n}_1}(x, t\to \infty)}\right)
%
%
\frac{u_{\bm{n}_i}(\bm{X}_{\bm{n}_i},
  t)}{\int_{-\infty}^{\infty}u_{\bm{n}_i}(\bm{X}_{\bm{n}_i}, t) \dd{X_{i,
      1}}}
\label{ubar}
\end{equation}
across the first 10 generations at $t=2$. Fig.~\ref{oscillatory}(a)
shows that division events, which bring newborn cells into later
generations $i \geq 2$, prevent structured cellular density in later
generations from reaching the equilibrium.

If the coefficients $g, \sigma, \beta, \tilde{\beta}$ depend only on
the cellular internal state $X$ and time $t$ and not on the cells'
generation, we can define
\begin{equation}
\hat{\rho}_{n}(\vec{X}_{n}, t) \coloneqq \sum_{\sum{n_i}=n}
\frac{1}{n!} \sum_{\pi} p_{\bm{n}}(\pi(\bm{X}_{\bm{n}}), t).
\label{hatrhodef}
\end{equation}
where $p_{\bm{n}}$ is defined in Eq.~\eqref{psdef} and the summation
over $\pi$ is over all possible rearrangements of $\bm{X}_n$ (defined
in Table~\ref{tab:model_variables}) of a generation-resolved cell
population $\bm{n}$ such that the union of states of all cells in all
generations is $\vec{X}_n$ (\textit{i.e.}, if we pad $\bm{X}_{\bm{n}}$
into one vector $(X_{1, 1}, X_{1, 2},..., X_{k, n_k})$, then such a
vector is a rearrangement of $\vec{X}_n$, the vector of attributes of
all $n$ cells as defined in Table~\ref{tab:model_variables}).


It can be shown that the differential equation satisfied by
$\hat{\rho}_n$ is

\begin{equation}
\begin{aligned}
  \f{\p \hat{\rho}_{n}}{\p t} + \sum_{j=1}^{n}\f{\p(g\hat{\rho}_{n})}{\p X_{ j}} = &
  \mfrac{1}{2} \sum_{j=1}^{n} \f{\p^2 (\sigma^2\hat{\rho}_{n})}{(\p X_{ j})^2}
  -\sum_{j=1}^{n}\big(\beta(X_{j})+\mu(X_{j})\big) \hat{\rho}_{n} \\
  \: & +\frac{1}{n}\sum_{j_1\neq j_2}\!\medint\int \tilde{\beta}(Y, X_{j_1}, X_{j_2})
  \hat{\rho}_{n-1}(\vec{X}_{n_{\text{b}}}^{{j_1, j_2}}, t)\dd{Y}
  + (n+1)\int\mu(Y)\hat{\rho}_{n+1}(\vec{X}_{n_{\text{d}}}, t)\, \dd{Y},
\end{aligned}
\end{equation}
where $\vec{X}_{n_{\text{b}}}^{j_1, j_2}$ is the pre-division cell
states that are different from $\vec{X}_n$ in that it does not have
contain the $X_{j_1}, X_{j_2}$ terms but has an extra $Y$ at the end;
$\vec{X}_{n_{\text{d}}}$ is the pre-death cell states that is
different from $\vec{X}$ in that it has an extra $Y$ component. In
this case, we can define the generation-independent marginalized
cell density

\begin{equation}
  u_n(\vec{X}_n, t) \coloneqq \sum_{m\geq n}(m)_{n}
  \int \hat{\rho}_{m}(\vec{X}_m, t)
  \dd \vec{X}_m\backslash \vec{X}_n
\label{un}
\end{equation}
which satisfies

\begin{equation}
\begin{aligned}
  \f{\p u_n(\vec{X}_n, t)}{\p t} + \sum_{j=1}^{n}\f{\p(gu_n)}{\p X_{j}} = & \mfrac{1}{2}
  \sum_{j=1}^{n} \f{\p^2 (\sigma^2u_n)}{(\p X_{j})^2} -\sum_{j=1}^{n}
  \big(\beta(X_{j})+\mu(X_{j})\big)u_n \\
  \: & + \sum_{j_1\neq j_2}\int \tilde{\beta}(Y, X_{j_1}, X_{j_2})
  u_{n-1}(\vec{X}_{n_{\text{b}}}^{{ j_1, j_2}}, t)\, \dd{Y} \\
  \: & + \sum_{j=1}^{n}\int\Big(\tilde{\beta}(Y, X_{ j}, Z)
  +\tilde{\beta}(Y, Z, X_{ j})\Big)u_{n}(\vec{X}_{{n}_{{\text{b}}}}^{{{ j}}}, t)\,\dd Y\dd Z.
\end{aligned}
\label{u_n}
\end{equation}
Here, $\vec{X}^{j}_{n_{\text{b}}}$ is different from $\vec{X}_n$ in
that $X_{j}$ is deleted, but an extra variable $Y$ is added as the
last component. If we take $n=1$, we can obtain a closed-form PDE for
describing the cell density w.r.t. the scalar state variable $X$

\begin{equation}
\begin{aligned}
  \f{\p u_1(X,t)}{\p t} +\f{\p(gu_1)}{\p X} = \mfrac{1}{2}
  \f{\p^2 (\sigma_j^2u_1)}{(\p X)^2}
  -\big(\beta(X)+\mu(X)\big)u_1 + \int\big(\tilde{\beta}(Y, X, Z)
+ \tilde{\beta}(Y, Z, X)\big)u_{1}(Y, t)\,\dd Y\dd Z.
\label{chpt9_v1}
\end{aligned}
\end{equation}
Eq.~\eqref{chpt9_v1} is equivalent to the cell sizer model, or a
timer-sizer model of cell division \cite{Xia_2021} after marginalizing
over the cells' ages.  As an implementation of this model, one can
numerically solve Eq. \eqref{u_n} or \eqref{chpt9_v1} using different
inferred single-cell-level gene expression dynamics as candidates for
$g$ \cite{Grima2022}.


\subsection{Evolution of cell numbers}
In the simplest case where all model parameters are constants, we can
marginalize over all cell state variables to obtain total cell
populations. More specifically, if we define the generation vector
$\bm{\bm{i}}\coloneqq(i_1,...,i_k), 0<i_1<...<i_k$ and the associated
orders of moments $\bm{\bm{\ell}}\coloneqq (\ell_1,...,\ell_k),
\ell_s>0$, then we can track the expectation of the product of
different orders of the number of cells in different generations

\begin{equation}
  \mathds{E}\Big[\prod_{s=1}^kn_{i_s}^{\ell_{s}}\Big]\coloneqq
  \sum_{\bm{n}}\prod_{s=1}^kn_{i_s}^{\ell_{s}}
  \int \rho_{\bm{n}}({\bm{X}_{\bm{n}}}, t)\,\dd{\bm{X}_{\bm{n}}}.
  \label{Erho}
\end{equation}
The differential equation satisfied by
$\mathds{E}\big[\prod_{i=1}^kn_i^{\ell_i}\big]$ can be shown to be

\begin{equation}
\begin{aligned}
  \frac{\dd\mathds{E}\big[\prod_{s=1}^{k}n_{i_s}^{\ell_s}\big]}{\dd t} = &
  \sum_{r=1, i_r>1}^k\beta_{i_r-1}\Big(\mathds{E}\big[\prod_{s=1}^k(n_{i_s}
    -\delta_{i_r-1, i_s}+2\delta_{i_r, i_s})^{\ell_s}
    n_{i_r-1}\big] - \mathds{E}\big[\prod_{s=1}^kn_{i_s}^{\ell_s} n_{i_r-1}\big] \Big)\\
  \: & +\sum_{r=1}^k\beta_{i_r}\Big(\mathds{E}\big[\prod_{s=1}^k
    (n_{i_s}-\delta_{i_r, i_s}+2\delta_{i_r+1, i_s})^{\ell_s}
    n_{i_r}\big] - \mathds{E}\big[\prod_{s=1}^kn_{i_s}^{\ell_s} n_{i_r}\big]\Big) \\
  \: & -\sum_{r=1}^{k-1} \beta_{i_r}\bigg(\delta_{i_{r+1}-i_r,1}
  \Big(\mathds{E}\big[\prod_{s=1}^k(n_{i_s}-\delta_{i_r, i_s}
    +2\delta_{i_r+1, i_s})^{\ell_s} n_{i_r}\big]
  - \mathds{E}\big[\prod_{s=1}^kn_{i_s}^{\ell_s} n_{i_r}\big]\Big)\bigg) \\
  \: & - \sum_{r=1}^{\infty}\mu_{i_r}
\Big(\mathds{E}\big[\prod_{s=1}^kn_{i_s}^{\ell_s} n_{i_r}\big]
  - \mathds{E}\big[\prod_{s=1}^k(n_{i_s}-\delta_{i_s, i_r})^{\ell_s} n_{i_r}\big]\Big).
\end{aligned}
\label{e_number}
\end{equation}
where $\delta_{i_{r}, i_{s}}=1$ if $i_{r}=i_{s}$ and
 $\delta_{i_{r}, i_{s}}=0$ otherwise is the Kronecker $\delta$-function. 
Note that if $\bm{i}=(i)$ is one-dimensional, and $\bm{\ell}=(1)$,
then Eq.~\eqref{e_number} reduces to the evolution of the average cell
number in the $i^{\text{th}}$ generation

\begin{equation}
\frac{\dd\mathds{E}[n_{i}]}{\dd t} = 2\beta_{i-1} \mathds{E}[n_{i-1}]
- \beta_{i}\mathds{E}[n_{i}] - \mu_{i} \mathds{E}[n_{i}].
\label{En_generations}
\end{equation}

Finally, we can consider another special simplifying case where

\begin{equation}
  P(\bm{n},t) \coloneqq \int \!\rho_{\bm{n}}(\bm{X}_{\bm{n}}, t)\dd{\bm{X}_{\bm{n}}}
  \label{P_generation}
\end{equation}
is the probability that the population contains $\{n_{1},
n_{2},\ldots, n_{k}$ cells in generations $1,\ldots, k$, respectively,
regardless of the individual's values of $X$.  It turns out that
$P(\bm{n})$ satisfies the  series of interdependent master equations
\begin{equation}
  \frac{\dd P(\bm{n},t)}{\dd t} = \sum_{i=1}^{k-1}
  \beta_{i-1}(n_{i-1}+1)P(\bm{n}_{\text{b}, i},t) 
-\sum_{i=1}^k \big(\beta_i +\mu_{i}\big) n_i P(\bm{n},t) +
  \sum_{i=1}^{\infty} \mu_i(n_i+1) P(\bm{n}_{\text{d},i},t).
  \label{P_generation_eq}
\end{equation}
when the division rates and the birth rates are constants within the
same generation, \textit{i.e.}, $\mu_{i, j}\equiv \mu_i, \beta_{i,
  j}\equiv \beta_i, \beta_0\coloneqq 0$. Eq.~\eqref{P_generation_eq}
is a multigenerational birth-death master equation for the number of
individuals in each generation $i$ which carries the same structure as
birth-death processes for cells grouped by different attributes other
than generation \cite{BDI}. Note that generating new members of a
successive generation arises only from birth, while death only
decreases the numbers within a generation.

\subsection{Evolution of ``biomass''}

Another quantity of specific interest is the biomass (\textit{e.g.},
the total amount of protein or mRNA within a subpopulation). For
example, the total mass within cells of the $i^{\rm th}$ generation
can be defined as $X_i \equiv \sum_{j=1}^{n_i}X_{i, j}$ and its
expectation evaluated from
%

\begin{equation}
  \mathds{E}\big[X_i(t)\big]
  = \sum_{\bm{n}}\int \Big(\sum_{j=1}^{n_i}X_{i, j}\Big)
  \rho_{\bm{n}}(\bm{X}_{\bm{n}}, t) \dd\bm{X}_{\bm{n}}
\label{volumeK}
\end{equation}
where $\rho_{\bm{n}}(\bm{X}_{\bm{n}}, t)$ is defined in Eq.~\eqref{symmeqn}.


In general, the differential equation satisfied by $X_i(t)$ involves
higher moment quantities; thus, the model is not closed.  However,
given certain constraints on the parameters, the dynamics for $X_i(t)$
can be closed, and a solution can be explicitly computed
(analytically, or numerically). For example, if $\beta_i(X_{i,
  j})\coloneqq \beta_i, \mu_i(X_{i, j})\coloneqq \mu_i$ are constants,
$g_i(X_{i, j})\coloneqq g_i X_{i, j}$ is linear, and the quantity $X$
is conserved across cell division (that is, if the mother cell carries
the state variable $X$ and the two daughter cells have are in states
$Y_1$ and $Y_2$, then $Y_1+Y_2=X$), then

\begin{equation}
  \frac{\dd \mathds{E}[X_i(t)]}{\dd t} = \big(g_i-\mu_{i}-\beta_{i}\big)
  \mathds{E}\big[X_i(t)\big] +\beta_{i-1} \mathds{E}\big[X_{i-1}(t)\big].
  \label{EX_i}
\end{equation}
Furthermore, if the growth rate and division rate are independent of
the generation number $i$, we can define expectations over any moment
of the total biomass summed over cells of all generations as

\begin{equation}
  \mathds{E}\big[X^q(t)\big]= \sum_{\bm{n}}\int \Big(\sum_{i=1}^{k}
  \sum_{j=1}^{n_i}X_{i, j}\Big)^q
  \rho_{\bm{n}}(\bm{X}_{\bm{n}}, t)\,\dd{\bm{X}_{\bm{n}}}, \quad q>1.
\label{Xhigher}
\end{equation}
Specifically, if $\mu$ is a constant and $g(X)= g_{0} X$ (and $X$ is
conserved across cell division), the differential equations satisfied
by the first and second moments of the \textit{total} biomass $X(t)$ and $X^2(t)$ are

\begin{equation}
\begin{aligned}
  \frac{\dd \mathds{E}[X(t)]}{\dd t} & = \big(g_0-\mu\big)\int x u_1(x, t)\dd x \equiv 
\big(g_{0} - \mu\big)
  \mathds{E}\big[X(t)\big],\\
%
\frac{\dd \mathds{E}\big[X^2(t)\big]}{\dd t} & = \big(g_{0}-2\mu\big)
\mathds{E}\big[X^2(t)\big]
+ \sigma^2 \mathds{E}\big[X(t)\big]
+ \mu\int x^{2} u_{1}(x,t)\dd x.
%
%
  \label{EXEX2}
\end{aligned}
\end{equation}
Only the equation for the mean total biomass $\mathds{E}[X(t)]$ is
closed. Its second moment depends on averages over $u_{1}(x,t)$
requiring the solution to Eq.~\eqref{chpt9_v1}.  General cases for the
equations satisfied by $\mathds{E}\big[X^q(t)\big]$ for arbitrary
$q\in\mathbb{N}^+$ are discussed in Appendix~\ref{appendixc}.

\subsection{Tracking dead cells}

Thusfar, we have assumed that the ``biomass'' $X$ originates from live
cells. Once cells die, they are no longer counted in the population
and the biomass $X$ associated with them is no longer included.
However, experimentally, the protein and/or mRNA extracted from a
solution of cells may come from both living and dead cells (at the
time of extraction). To describe these types of measurements, we
keep track of cells that have died and assign them to the
$0^{\text{th}}$ generation $g_{0} = \beta_0 =0$. We denote their
states by $\bm{X}_0\coloneqq (X_{0, 1},...,X_{0, n_0})$. We then
define $\tilde{p}_{\bm{n}}(\bm{X}_{\bm{n}},t|\bm{X}(0)_{\bm{n}(0)},
0)$ to include the zero-generation (cells that have died)
population. Using arguments similar to those in Proposition
\ref{Markovjump} we can show that under certain conditions
$\tilde{p}_{\bm{n}}$ satisfies the differential equation
\begin{equation}
\begin{aligned}
  \f{\p \tilde{p}_{\bm{n}}}{\p t} + \sum_{i=1}^k\sum_{j=1}^{n_i}
  \f{\p(g_{i, j}\tilde{p}_{\bm{n}})}{\p X_{i, j}} = & \mfrac{1}{2} \sum_{i=1}^k\sum_{j=1}^{n_i}
  \f{\p^2 (\sigma^2_{i, j}\tilde{p}_{\bm{n}})}{(\p X_{i, j})^2} - \sum_{i=1}^k
  \sum_{j=1}^{n_i}\big(\beta_{i, j}+\mu_{i, j}\big) \tilde{p}_{\bm{n}} \\
  \: & \,\, + \sum_{i=1}^{k-1}\sum_{j=1}^{n_{i-1}+1}\!nt
  \tilde{\beta}_{i,j}(Y, X_{i+1, n_{i+1}-1}, X_{i+1, n_{i+1}})
  \tilde{p}_{\bm{n}_{\text{b}, i}}(\bm{X}_{\bm{n}_{\text{b}, i}^{j}}\!t \vert
  \bm{X}(0)_{\bm{n}(0)},0)\,\dd{Y} \\
  \: & \,\, + \sum_{i=1}^{\infty}\sum_{j=1}^{n_i+1}\mu(X_{0, n_0})
  \tilde{p}_{\bm{n}_{\tilde{\text{d}}, i}}(\bm{X}_{\bm{n}_{\tilde{\text{d}}, i}}^j\!
  t|\bm{X}(0)_{\bm{n}(0)}, 0)
\end{aligned}
\end{equation}
where $\bm{n}_{\tilde{\text{d}}, i}$ differs from in that its
$0^{\text{th}}$ component is $n_0-1$ but its $i^{\text{th}}$ component
is $n_i+1$, and $\bm{X}_{\bm{n}_{\tilde{\text{d}}, i}}^j$ differs from
$\bm{X}_{\bm{n}}$ in that the internal states of the $0^{\text{th}}$
generation (dead cells) are $(X_{0, 1},..., X_{0, n_0-1})$ and the
internal states of the $i^{\text{th}}$ generation are $(X_{i,
  1},...,X_{i, j-1}, X_{0, n_0}, X_{i, j},...,X_{i, n_i })$ ($X_{0,
  n_0}$ is in the $j^{\text{th}}$ component). Similarly, we can define
the unconditional probability density function
$\tilde{p}_{\bm{n}}^{*}(\bm{X}_{\bm{n}}, t)$ as defined in
Eq.~\eqref{psdef} as well as the symmetrized probability density
function
\begin{equation}
  \tilde{\rho}_{\bm{n}}(\bm{X}_{\bm{n}}, t)
  \coloneqq \prod_{i=0}^k \frac{1}{n_i!} \sum_{\pi}
  \tilde{p}_{\bm{n}}^{*}(\pi(\bm{X}_{\bm{n}}), t).
\label{rhodef1}
\end{equation}
The PDE satisfied by $\tilde{\rho}_{\bm{n}}$ is
\begin{equation}
\begin{aligned}
 \f{\p \tilde{\rho}_{\bm{n}}}{\p t} + \sum_{i=1}^k\sum_{j=1}^{n_i}
 \f{\p(g_{i, j} \tilde{\rho}_{\bm{n}})}{\p X_{i, j}} = &  \mfrac{1}{2} \sum_{i=1}^k\sum_{j=1}^{n_i}
 \f{\p^2 (\sigma_{i, j}^2\tilde{\rho}_{\bm{n}})}{(\p X_{i, j})^2} -
 \sum_{i=1}^k\sum_{j=1}^{n_i}\big(\beta_{i, j}+\mu_{i, j}\big) \tilde{\rho}_{\bm{n}} \\
 \: & \,\, +\sum_{i=1}^{k-1}\frac{n_{i}+1}{n_{i+1}(n_{i+1}-1)}
 \sum_{j_1\neq j_2}\!\medint\int \!\tilde{\beta}_{i, j}(Y, X_{i+1, j_1}, X_{i+1, j_2})
 \tilde{\rho}_{\bm{n}_{\text{b}, i}}(\bm{X}_{\bm{n}_{\text{b}, i}}^{{j_1, j_2}}, t)\,\dd{Y} \\
 \: &\,\, + \frac{1}{n_0}\sum_{i=1}^{\infty}(n_i+1)\sum_{j=1}^{n_0}
 \mu_{i, j}(X_{0, j})\rho_{\bm{n}_{\tilde{\text{d}}, i}}(\bm{X}_{\bm{n}_{\tilde{\text{d}}, i}}^j, t).
 \label{tilde_rho}
\end{aligned}
\end{equation}

The expectation of the total biomass $X_{0}\equiv \sum_{j=1}^{n_0}X_{0, j}$
associated with dead cells can be found from
\begin{equation}
  \mathds{E}\big[X_{0}(t)\big] \equiv
  \sum_{\bm{n}}\int \big(\sum_{j=1}^{n_0}X_{0, j}\big)
  \tilde{\rho}_{\bm{n}}(\bm{X}_{\bm{n}}, t)\,\dd {\bm{X}_{\bm{n}}}.
\end{equation}
If the death rates $\mu_{i}$ of cells are equal and constant within
each generation $i$, then  $\mathds{E}\big[X_{0}(t)\big]$ satisfies
\begin{equation}
  \frac{\dd  \mathds{E}\big[X_{0}(t)\big]}{\dd t}
  = \sum_{i=1}^{\infty} \mu_i  \mathds{E}\big[X_i(t)\big],
  \label{EX0}
\end{equation}
where $ \mathds{E}\big[X_i(t)\big]$ is the total expected biomass from
  cells in the $i^{\text{th}}$ generation, as defined in
  Eq.~\eqref{volumeK}.

We can also define second moments involving the biomass from dead cells
\begin{equation}
\begin{aligned}
   \mathds{E}\big[X_{0}^{2}(t)\big] = \sum_{\bm{n}}\int \big(\sum_{j=1}^{n_0}X_{0, j}\big)^2
   \tilde{\rho}_{\bm{n}}(\bm{X}_{\bm{n}}, t)\,\dd{\bm{X}_{\bm{n}}}
   \label{EX020}
\end{aligned}
\end{equation}
and
\begin{equation}
  \mathds{E}\big[X_{0}(t)X(t)\big] = \sum_{\bm{n}}\int
  \Big(\sum_{i=1}^{k}\sum_{j=1}^{n_i}X_{i, j}\Big)
   \Big(\sum_{\ell=1}^{n_0}X_{0, \ell}\Big)
   \tilde{\rho}_{\bm{n}}(\bm{X}_{\bm{n}}, t)\,\dd {\bm{X}_{\bm{n}}}.
    \label{EX0EX}
\end{equation}
If we assume that the death rate is a constant $\mu$ for all cells,
the growth rate $g(X) = g_{0}X$, and the state variable $X$ is
conserved at division, we can derive the differential equations

\begin{equation}
  \frac{\dd  \mathds{E}\big[X_{0}^{2}(t)\big]}{\dd t}
  = 2\mu  \mathds{E}\big[X_{0}(t)
    X(t)\big]
+ \mu \sum_{\bm{n}}\int\Big(\sum_{i=1}^{k}\sum_{j=1}^{n_i}X_{i, j}^2\Big)
\tilde{\rho}_{\bm{n}}(\bm{X}_{\bm{n}}, t)\,\dd \bm{X}_{\bm{n}}
\label{EX02}
\end{equation}
%
%
\begin{equation}
  \frac{\dd \mathds{E}\big[X_{0}(t)X(t)\big]}{\dd t} = (g_{0}-\mu)
  \mathds{E}\big[X_{0}(t)X(t)\big] +
  \mu  \mathds{E}\big[X^2(t)\big]- \mu \sum_{\bm{n}}
\int \Big(\sum_{i=1}^{k}\sum_{j=1}^{n_i}X_{i, j}^2\Big)
\tilde{\rho}_{\bm{n}}(\bm{X}_{\bm{n}}, t)\,\dd{\bm{X}_{\bm{n}}}.
%
%
  \label{EX0X}
\end{equation}
Higher moments of $X_{0}, X$ can also be evaluated, which we do not
include for brevity.
%

\subsection{Correlations and interactions}

Although examples so far have involved simple forms of $g, \sigma,
\beta, \mu$ that depend only on the state of of the cell being
tracked, these rates can depend on the states of other cells in the
population.  These more complex dependences prevent closure of the
PDEs and signal more complex correlations, or ``interactions.''
Simple interactions can be incorporated in the ``mean-field'' limit if
we consider the parameters $g, \sigma, \beta, \mu$ to be functions of
only averaged macroscopic quantities such as $X(t)$.

As an intuitive example, if we allow the death rate of the
$j^{\text{th}}$ cell in the $i^{\text{th}}$ generation to also depend
on the total ``biomass'' from all living cells, $\mu_{i, j} = \mu_{i,
  j}(X_{i, j}, \sum_{i}\sum_{j=1}^{n_i} X_{i, j})$.  Using this form of
death rate leads to a symmetric population density
$\rho_{\bm{n}}(\{X\}_{\bm{n}}$ that satisfies

\begin{equation}
\begin{aligned}
  \f{\p \rho_{\bm{n}}}{\p t} + \sum_{i=1}^k\sum_{j=1}^{n_i}
  \f{\p(g_{i}\rho_{\bm{n}})}{\p X_{i, j}}
  = &  \mfrac{1}{2} \sum_{i=1}^k\sum_{j=1}^{n_i}
  \f{\p^2 (\sigma_{i, j}\rho_{\bm{n}})}{(\p X_{i, j})^2}
  -\sum_{i=1}^k\sum_{j=1}^{n_i}\Big(\beta_{i, j}(X_{i, j})+\mu_{i, j}\big(X_{i, j},
  {\displaystyle\medop\sum_{\ell=1}^k
    \medop\sum_{m=1}^{n_{\ell}}X_{\ell, m}}\big)\Big) \rho_{\bm{n}} \\
  \:  & \quad +\sum_{i=1}^{k-1}\frac{n_{i}+1}{n_{i+1}(n_{i+1}-1)}
  \!\sum_{1\leq j_1\neq j_2\leq n_{i+1}}\!\!\!\!\!int\int
  \tilde{\beta}_{i}(Y, X_{i+1, j_1}, X_{i+1, j_2})
  \rho_{\bm{n}_{\text{b}, i}}(\bm{X}_{\bm{n}_{\text{b}, i}}^{{j_1, j_2}}, t)\,\dd{Y} \\
  \: & \quad + \sum_{i=1}^{\infty}\sum_{j=1}^{n_i+1}\int
  \mu_{i, j}\big(Y,\, {\displaystyle \medop\sum_{\ell=1}^k
    \medop\sum_{m=1}^{n_{\ell}}X_{\ell, m}+Y}\big)
  \rho_{\bm{n}_{\text{d}}}(\bm{X}_{\bm{n}_{\text{d}, i}}^{{ j}}, t) \dd{Y}.
\end{aligned}
\label{symmeqn1}
\end{equation}
Due to the dependencies on the mean-field term
$\sum_{i=1}^k\sum_{j=1}^{n_i} X_{i, j}$, we cannot obtain a
closed-form equation for macroscopic quantities such as the cellular
density $u_{1}(X_1, t)$ defined in Eq.~\eqref{chpt9_v1}. However, if
the approximation $\sum_{i=1}^k\sum_{j=1}^{n_i}X_{i, j}\approx
\sum_{i=1}^k\sum_{j=1}^{n_i}X_{i, j}+Y \approx \mathds{E}[X(t)]$ can
be made, with $\mathds{E}[X(t)]$ defined in Eq.~\eqref{volumeK}, an
approximate PDE for $u_{1}(X, t)$ defined in Eq.~\eqref{un} can be
motivated:

\begin{equation}
\begin{aligned}
  \f{\p u_1(X_{1}, t)}{\p t} +\f{\p(gu_1)}{\p X_1} = & 
  \mfrac{1}{2} \f{\p^2 (\sigma_j^2u_1)}{(\p X_1)^2}
  -\Big(\beta(X_1)+\mu\big(X_1,\medint\int Y u_1(Y, t)\text{d}Y \big)\Big) u_1(X, t) \\
  \: & \qquad\qquad\qquad  +\! \int\!\Big(\tilde{\beta}(Y, X_1, Z)
+ \tilde{\beta}(Y, Z, X_1)\Big)u_{1}(Y, t)\,\dd Y\dd Z.
\end{aligned}
\label{u_structured1}
\end{equation}
Eq.~\eqref{u_structured1} is nonlinear because the mean-field term
depends on $\int x u_1(x, t)\text{d}x$. Similarly, if other
coefficients depend on mean-field quantities or some specific
interaction terms among cells exist, then by making assumptions and
marginalization, it might still be possible to find self-consistent
integrodifferential equations for macroscopic quantities of
interest. For example, death rates that depend on the values of $X$ of
two different cells have been shown to generate a nonlinear
interaction term in kinetic derivations of single-species
predator-prey type models \cite{xia2023population}.

\section{Summary and Conclusions}
\label{conclusion}
In this work, we used the forward-type Feynman-Kac formula and Markov
jump process to formulate a kinetic theory for describing the cellular
population density of a generation-resolved cellular population with
fluctuating rates of changing internal states as well as random
division times. Such a general kinetic theory not only tracks each
cell's continuous-valued state attribute such as its volume, protein
or mRNA abundance, but also its generation (\textit{i.e.}, how many
times its ancestors have divided). In general, our kinetic theory
framework can apply to any collection of particles that experience
demographic noise from birth-death processes as well as noise in
specific individual-level attributes.

A number of new results were presented.  The underlying kinetic theory
describing the intra-generation-symmetrized cell populations is given
by Eq.~\eqref{symmeqn} (or Eq.~\eqref{symmeqn_vector} for a vector of
attributes). We find that this fully resolved, high-dimensional
probability density can be marginalized in to different
directions. First, one can sum over moments of the discrete
populations/subpopulations to find the dynamics of a generalized cell
population density $u_{\bm{n}}(\bm{X}_{\bm{n}},t)$
(Eq.~\eqref{u_definition}), which is found to obey
Eq.~\eqref{u_structured} when generations are tracked, and
Eq.~\eqref{chpt9_v1} in the generation-independent case. Further
marginalizing over all cell attributes $\bm{X}_{\bm{n}}$ allows one to
derive simpler equations for useful quantities such as the expected
total number of cells in each generation (Eq.~\eqref{En_generations})
and the generation structure of the total population
(Eq.~\eqref{P_generation_eq}).

Alternatively, the full probability densities can be used to define
moments of mean-field quantities such as total gene expression levels
or biomass $X$ across the entire population. These are derived in
Eqs.~\eqref{EX_i} and \eqref{EXEX2}, which depend on integrals over
the single-particle number density $u_{1}(x,t)$.  We also show how the
biomass $X_{0}$ from dead calls can also be tracked, as is often the
case in experiments. Expressions for the lowest moments are given in
Eqs.~\eqref{EX0}, \eqref{EX02}, and \eqref{EX0X}. Our results are
tabulated below:

\begin{table}[h!]
\centering
\renewcommand*{\arraystretch}{1.1}
\begin{tabular}{| >{\centering\arraybackslash} m{11em}| 
    >{\arraybackslash} m{20em}| >{\arraybackslash} m{17em}|}
  \hline \textbf{Quantity} &
  \textbf{Meaning} &  \textbf{Equation}\\[1pt]
  \hline\hline
  \,\, $u_{\bm{n}}(\bm{X}_{\bm{n}},t)$\,\, &  partially marginalized cell
  population density of any order &
  Eq.~\eqref{u_structured}. Closed set of
  PDEs for noninteracting systems \\[2pt] \hline
  \,\, $u_n(\vec{X}_{n},t)$\,\, & generation-independent cell
  population density (may include intercellular dependence) & Eqs.~\eqref{un} and \eqref{u_n}
  \\[2pt]
  \hline
  \,\, $\mathbb{E}[\bm{n}(t)]$\,\, & expectation of moments of total cell number &
  Eqs.~\eqref{Erho}, \eqref{e_number}, and \eqref{En_generations}
  \\[2pt] \hline
  \,\, $P(\bm{n},t)$\,\, & probability of $\bm{n}=\{n_{i}\}$ in each generation $i$ &
  Eqs.~\eqref{P_generation} and \eqref{P_generation_eq}
  \\[2pt] \hline
  \,\, $\mathbb{E}[X^q(t)]$\,\, &
  expectation of moments of total biomass or expression levels
  & Eqs.~\eqref{Xhigher}, \eqref{EXEX2}, and \eqref{chpt9_v1} \\[2pt]
  \hline
  \,\, $\mathbb{E}[X_0^{p}(t)X^{q}(t)], \, p+q\leq 2$\,\, & mean and variance of
  biomass from dead cells & Eqs.~\eqref{EX0},\eqref{EX020},\eqref{EX0EX},
  \eqref{EX02}, \eqref{EX0X}, and \eqref{tilde_rho}\\[2pt]
  \hline
\end{tabular}
\vspace{1mm}
\caption{\small\textbf{Summary of our main results.} Functions
  describing cell numbers and overall attributes are listed, along
  with the equation numbers of their mathematical definitions and
  dynamical equations.}
\label{tab:results}
\end{table}

Finally, we discussed cell-cell ``interactions'' that manifest
themselves through birth or death rates that depend on the attribute
$X_{i,j}$ of multiple cells.  Such forms of the birth and death rate
precludes full marginalization, leading to higher order correlation
terms for which an approximation must be imposed to close the
equations. We show how a death rate that also depends on the total
biomass results in the implicitly nonlinear (in $u_{1}(x,t)$) PDE
given in ~\eqref{u_structured1}.



\added{Note that the PDEs for marginalized densities
  $u_{\bm{n}}(\bm{X}_{\bm{n}},t)$ can be solved numerically using
  newly developed adaptive spectral methods suited for unbounded
  domains \cite{xia2021efficient,xia2021frequency,chou2023adaptive}
  and provide an ``Eulerian'' representation of the structured
  population.  Our kinetic theory/PDE framework does not directly
  track the structure of populations along lineages of cells (a more
  ``Lagrangian'' picture) but connecting our Eulerian representation
  with representations that delineate cell lineages would be useful
  area of future analysis.}

Other directions for future analysis include developing tractable
models of interactions that arise through complex dependences of birth
and death rates on $\bm{X}_{\bm{n}}$. \added{The equations we derived
  can also inform inverse-type problems by serving as constraints for
  neural network-based machine learning approaches for inferring model
  parameters (such as interacting birth and death rates) from data
  \cite{xia2022spectrally}. Structured populations that vary spatially
  also arise in many applications
  \cite{ayati2007,auger2008structured,nadell2016spatial}.  For models
  in which convection and diffusion depend on expression levels, the
  dynamics of $\bm{X}_{\bm{n}}$ can be modeled as being coupled to
  transport.}


Data Availability The datasets generated during and/or analysed during the current study are available from
the corresponding author on reasonable request.
Declarations
Conflict of Interest The authors declare that they have no known competing financial interests or personal
relationships that could have appeared to influence the work reported in this paper.

\bibliography{bibliography}

\newpage

\appendix

\section{Derivation of the differential equation satisfied by
  the cell population probability density function}
\label{derivationsde}
\setcounter{equation}{0}
\renewcommand{\theequation}{A\arabic{equation}}

To show $p_{\bm{n}}(\bm{X}_{\bm{n}}, t|\bm{X}(0)_{\bm{n}(0)}, 0)$
defined in Eq.~\eqref{pdef} satisfies Eq.~\eqref{ppde2}, we require
the following two propositions.\\

\begin{prop}
\label{lemma1}
\rm (Forward-type Feynman-Kac formula) If the coefficients $g_{i, j},
\sigma_{i, j}, \beta_{i, j}, \mu_{i, j}$ are smooth, uniform Lipschitz
continuous, and uniformly bounded, then, under certain conditions,
the solution to the following PDE

\begin{equation}
\begin{aligned}
  \f{\p \hat{p}_{\bm{n}}(\bm{X}_{\bm{n}}, t|\bm{X}(0)_{\bm{n}(0)}, 0)}{\p t}
  +\! \sum_{i=1}^k\sum_{j=1}^{n_i}\f{\p(g_{i, j}(X_{i, j}, t)\hat{p}_{\bm{n}})}{\p X_{i, j}} = & 
\mfrac{1}{2} \sum_{i=1}^k\sum_{j=1}^{n_i} \f{\p^2 (\sigma_{i, j}^2(X_{i, j}, t)
    \hat{p}_{\bm{n}})}{(\p X_{i, j})^2} \\
\: & \quad  -\sum_{i=1}^k\sum_{j=1}^{n_i}\!\big(\beta_{i, j}(X_{i, j})
 +\mu_{i, j}(X_{i, j})\big) \hat{p}_{\bm{n}}
\end{aligned}
\label{forwardFeymann}
\end{equation}
with initial condition $\hat{p}_{\bm{n}}(\bm{X}_{\bm{n}}, 0) =
\delta(\bm{X}(0)_{\bm{n}(0)}-\bm{X}_{\bm{n}})$ if $\bm{n}=\bm{n}(0)$
(and $\hat{p}_{\bm{n}}(\bm{X}(0)_{\bm{n}(0)},0) =0$ if
$\bm{n}\neq\bm{n}(0)$), is
\begin{equation}
\begin{aligned}
  \hat{p}_{\bm{n}}(\bm{X}_{\bm{n}}, t|\bm{X}(0)_{\bm{n}(0)}, 0)
  \coloneqq\mathbb{E}\Big[\delta(\bm{X}(t)_{\bm{n}(t)}
    - \bm{X}_{\bm{n}})S\big(t; \bm{X}(t)_{\bm{n}(t)}\big)
\Big|\bm{X}(0)_{\bm{n}(0)}, 0;\bm{n}(0\leq s \leq t)=\bm{n}(0)\Big]
\end{aligned}
\end{equation}
%
%
where each component in $\bm{X}(t)_{\bm{n}(t)}$ satisfies Eq.~\eqref{Xevolve}.
\end{prop}

Proposition \ref{lemma1} provides the PDE satisfied by the density
function for all cells with states $\bm{X}_{\bm{n}}$ in the absence of
division and death. The proof of Proposition~\ref{lemma1} and the
associated specific technical assumptions are given in
section~\ref{appendixA} below.

When cell division or death occurs, the total number of cells changes
according to a Markov jump process. Thus, we need the following
proposition to derive the differential equation satisfied by the
conditional probability density function $p_{\bm{n}}(\bm{X}_{\bm{n}},
t|\bm{X}(0)_{\bm{n}(0)}, 0)$ defined in Eq.~\eqref{pdef}.\\

\begin{prop}
\label{Markovjump}
\rm (Markov jump process) Given the initial condition $\bm{n}(0)$ with
states $\bm{X}(0)_{\bm{n}(0)}$ at $t=0$ and a target state at time $t$
with $\bm{n}$ cells and their internal states $\bm{X}_{\bm{n}}$, we
start with the conditions

\begin{equation}
\begin{aligned}
p^{0}_{\bm{n}}(\bm{X}_{\bm{n}}, t|\bm{X}(0)_{\bm{n}(0)}, 0) \coloneqq & 0,\\
p^{1}_{\bm{n}}(\bm{X}_{\bm{n}}, t|\bm{X}(0)_{\bm{n}(0)}, 0)
  \coloneqq &\hat{p}_{\bm{n}}(\bm{X}_{\bm{n}}, t|\bm{X}(0)_{\bm{n}(0)}, 0),
\end{aligned}
\end{equation}
and recursively define

\begin{equation}
\begin{aligned}
p_{\bm{n}}^{m+1}(\bm{X}_{\bm{n}}, t|\bm{X}(0)_{\bm{n}(0)}, 0)
\coloneqq & \hat{p}_{\bm{n}}(\bm{X}_{\bm{n}}, t|\bm{X}(0)_{\bm{n}(0)}, 0)\\
\: & 
+\int_0^t \mathbb{E}\Big[
    S\big(\tau; \bm{X}(\tau)_{\bm{n}(\tau)}\big)
J^m\big( t,\tau;\bm{X}_{\bm{n}},\bm{n}(0)\big)
 \Big|\bm{X}(0)_{\bm{n}(0)}, 0; \bm{n}(0<s<\tau)=\bm{n}(0) \Big]\dd{\tau},
\end{aligned}
\label{recursion}
\end{equation}
where the birth-death probability flux is defined by

\begin{equation}
\begin{aligned}
J^m\big(t,\tau; \bm{X}_{\bm{n}}, \bm{n}(0)\big)\coloneqq & 
\sum_{i=1}^{k(0)}\sum_{j=1}^{n_{i}(0)}\Big[\tilde{\beta}_{i, j}
\big(X_{i, j}(\tau), X_1(\tau), X_2(\tau)\big)
p_{\bm{n}}^m(\bm{X}_{\bm{n}}, t-\tau|\bm{X}(\tau)_{\bm{n}(0)_{\text{b}, -i}}^{-j}, 0) \\
\: & \hspace{3.6cm} + \mu_{i, j}\big(X_{i, j}(\tau)\big)
p_{\bm{n}}^m(\bm{X}_{\bm{n}}, t-\tau|
\bm{X}(\tau)_{\bm{n}(0)_{ \text{d}, -i}}^{-j}, 0)\Big].
\label{Jm}
\end{aligned}
\end{equation}
Then, $p_{\bm{n}}^{m+1}$ satisfies

\begin{equation}
\begin{aligned}
  \f{\p p_{\bm{n}}^{m+1}}{\p t} + \sum_{i=1}^k\sum_{j=1}^{n_i}
  \f{\p(g_{i, j}p^{m+1}_{\bm{n}})}{\p X_{i, j}}
  = &   \mfrac{1}{2} \sum_{i=1}^k\sum_{j=1}^{n_i}
  \f{\p^2 (\sigma_{i, j}^2p^{m+1}_{\bm{n}})}{(\p X_{i, j})^2}
  - \sum_{i=1}^k\sum_{j=1}^{n_i}\big(\beta_{i, j}(X_{i, j})+\mu_{i, j}(X_{i, j})\big)
  p^{m+1}_{\bm{n}}\\
  \: & + \sum_{i=1}^{k-1}\sum_{j=1}^{n^{\text{b}}_{i-1}}\!\int
  \tilde{\beta}(Y, X_{i+1, n_{i+1}-1}, X_{i+1, n_{i+1}})
  p^{m}_{\bm{n}_{\text{b}, i}}(\bm{X}^j_{\bm{n}_{\text{b}, i}}, t\,\vert\,
  \bm{X}(0)_{\bm{n}(0)}, 0)\dd{Y} \\
  \: & + \sum_{i=1}^{\infty}\sum_{j=1}^{n_i^{\text{d}}}\int\!\mu(\Y)
  p^{m}_{\bm{n}_{\text{d}, i}}(\bm{X}^{j}_{\bm{n}_{\text{d}, i}}, t\, \vert\,
  \bm{X}(0)_{\bm{n}(0)}, 0) \dd{\Y}.
\end{aligned}
\label{induct}
\end{equation}
%
%
Furthermore, $p_{\bm{n}}^{m}$ is non-decreasing in $m$.
\end{prop}

The proof of Proposition~\ref{Markovjump} will be given in
section~\ref{appendixB} below. Intuitively, $m$ in
Eq.~\eqref{recursion} is the maximal number of birth or death events
allowed within the cell population. Since $p_{\bm{n}}^{m}$ is
increasing in $m$, there exists a $p^*$ such that $p^{m}\rightarrow
p^*$ \text{a.s.} for all $ \bm{X}(0)_{\bm{n}(0)}$ and
$\bm{X}_{\bm{n}}$. After integrating over $\bm{X}_{\bm{n}}$ and
summing over all $\bm{n}$ on both sides of Eq.~\eqref{recursion} and
assuming

\begin{equation}
  \sum_{{\bm{n}}}\int p_{\bm{n}}^{m-1}(\bm{X}_{\bm{n}}, t|\bm{X}(0)_{\bm{n}(0)}, 0)\,
  \dd{\bm{X}_{\bm{n}}}\leq 1
\label{induct_step}
\end{equation}
for $m\in\mathbb{N}$ and any initial condition
$\bm{X}(0)_{\bm{n}(0)}$, we have $\sum_{{\bm{n}}}\int
p_{\bm{n}}^{m}(\bm{X}_{\bm{n}}, t\,\vert\,\bm{X}(0)_{\bm{n}(0)}, 0)
\dd{\bm{X}_{\bm{n}}}\leq F^{m}(t; \bm{X}(0)_{\bm{n}(0)}, 0)$ where

\begin{equation}
  {\small
    \begin{aligned}
  F^{m}(t; \bm{X}(0)_{\bm{n}(0)}, 0) \coloneqq & \!\int
  \hat{p}_{\bm{n}(0)}(\bm{X}_{\bm{n}(0)}, t\,\vert\,\bm{X}(0)_{\bm{n}(0)}, 0) \dd\bm{X}_{\bm{n}(0)} \\
  \: &+ \int_0^t \mathbb{E}\Big[S\big(\tau; \bm{X}(\tau)_{\bm{n}(\tau)}\big)
   \sum_{i=1}^{k(0)}\sum_{j=1}^{n_i(0)} \big(\beta(X_{i, j}(\tau))+ \mu(X_{i, j}(\tau)\big)\big|
   \bm{X}(0)_{\bm{n}(0)}, 0;\bm{n}(0<s<\tau)=\bm{n}(0)\Big]\,\dd{\tau}
\end{aligned}}
\label{F_definition}
\end{equation}
and $S\big(\tau; \bm{X}(\tau)_{\bm{n}(\tau)}\big)$ is defined in
Eq.~\eqref{Wdef}.
%
%
Taking the derivative of $F^{m}(t; \bm{X}(0)_{\bm{n}(0)}, 0)$, we find
$\dd F^{m}(t; \bm{X}(0)_{\bm{n}(0)}, 0)/\dd t=0$.
It is straightforward to verify that $F^{m}(0; \bm{X}(0)_{\bm{n}(0)},
0)=1$; therefore, we have $F^{m}(t; \bm{X}(0)_{\bm{n}(0)}, 0) \equiv 1,
\forall t\geq 0$, which indicates that

\begin{equation}
\sum_{\bm{n}}\int p_{\bm{n}}^{m}(\bm{X}_{\bm{n}}, t|\bm{X}(0)_{\bm{n}(0)}, 0) 
\dd\bm{X}_{\bm{n}}\leq 1. 
\end{equation}
By induction, Eq.~\eqref{induct_step} holds true for all
$m\in\mathbb{N}^+$. Finally, it is easy to show that
$p_{\bm{n}}^{m}(\bm{X}(0)_{\bm{n}}, t|\bm{X}(0)_{\bm{n}(0)}, 0)\geq0$,
so by the monotone convergence theorem,
\begin{equation}
\sum_{\bm{n}}\int p_{\bm{n}}^{*}(\bm{X}_{\bm{n}}, t|\bm{X}(0)_{\bm{n}(0)}, 0) 
\dd\bm{X}_{\bm{n}}\leq 1,
\end{equation}
which indicates $0\leq p^*<\infty$ exists $\text{a.e.}$.

If i) the convergence $p^{m}\rightarrow p^*$ is uniform and ii) taking
the limit w.r.t. $m$ is interchangeable with taking the partial
derivatives in Eq.~\eqref{induct}, then $p^*$ is the solution to
\begin{equation}
\begin{aligned}
  \f{\p p_{\bm{n}}^{*}}{\p t} + \sum_{i=1}^k\sum_{j=1}^{n_i}
  \f{\p(g_{i, j}p^{*}_{\bm{n}})}{\p X_{i, j}} = &
  \mfrac{1}{2} \sum_{i=1}^k\sum_{j=1}^{n_i}
  \f{\p^2 (\sigma_{i, j}p^{*}_{\bm{n}})}{(\p X_{i, j})^2}
  -\sum_{i=1}^k\sum_{j=1}^{n_i}\big(\beta(X_{i, j})+\mu(X_{i, j})\big) p^{*}_{\bm{n}}\\
  \: & + \sum_{i=1}^{k-1}\sum_{j=1}^{n_{i}+1}\!\int
  \tilde{\beta}_{i, j}(Y, X_{i+1, n_{i+1}-1}, X_{i+1, n_{i+1}})
  p^{*}_{\bm{n}_{\text{b}, i}}(\bm{X}^j_{\bm{n}_{\text{b}, i}}, t|\bm{X}(0)_{\bm{n}(0)}, 0)\dd{Y}\\
  \: & + \sum_{i=1}^{\infty}\sum_{j=1}^{n_i+1}\!\int\mu_{i, j}(\Y)
  p^{*}_{\bm{n}_{\text{d}, i}}(\bm{X}^j_{\bm{n}_{\text{d}, i}}, t|\bm{X}(0)_{\bm{n}(0)}, 0) \dd{\Y}
\end{aligned}
\end{equation}
Since by taking the limit $m\rightarrow\infty$ in
Eq.~\eqref{recursion}, $p^*$ can also be written as
%
\begin{equation}
\begin{aligned}
  p^*_{\bm{n}}(\bm{X}_{\bm{n}}, t|\bm{X}(0)_{\bm{n}(0)}, 0) = & 
\hat{p}_{\bm{n}}(\bm{X}_{\bm{n}}, t|\bm{X}(0)_{\bm{n}(0)}, 0) \\
  \: & +\int_0^t \mathbb{E}\Big[
    S\big(\tau; \bm{X}(\tau)_{\bm{n}(\tau)}\big)J^{m}(t,\tau;\bm{X}_{\bm{n}},\bm{n}(0))
\Big|\bm{X}(0)_{\bm{n}(0)}(0), 0; \bm{n}(0<s<\tau)=\bm{n}(0)\Big]\, \dd{\tau}
\end{aligned}
\label{psformula}
\end{equation}
which solves the differential equation Eq.~\eqref{ppde2}. 

Finally, the definition of $p^*$ in Eq.~\eqref{psformula} coincides
with the definition of $p$ in Eq.~\eqref{pdef}. Thus, if
Eq.~\eqref{psformula} defined a unique $p^*$, then
$p^*_{\bm{n}}(\bm{X}_{\bm{n}}, t|\bm{X}(0)_{\bm{n}(0)},
0)=p_{\bm{n}}(\bm{X}_{\bm{n}}, t|\bm{X}(0)_{\bm{n}(0)},
0)$. Therefore, $p_{\bm{n}}$ also solves the differential equation
Eq.~\eqref{ppde2}. Specifically, if
\begin{equation}
\sum_{\bm{n}}\int p_{\bm{n}}(\bm{X}_{\bm{n}}, t|\bm{X}(0)_{\bm{n}(0)}, 0) \dd\bm{X}_{\bm{n}}=1,
\end{equation}
then $p_{\bm{n}}$ is indeed a probability density function of the
total cell population that satisfies Eq.~\eqref{ppde2}.

\subsection{Proof of Proposition~\ref{lemma1}}
\label{appendixA}
Here, we prove Proposition~\ref{lemma1} and provide the needed
technical assumptions. We shall apply Theorem 6.2 in
\cite{lecavil2015probabilistic}. If $\bm{n}\neq\bm{n}(0)$, then by
definition $\hat{p}_{\bm{n}} = 0$, which solves
Eq.~\eqref{forwardFeymann}. If $\bm{n}(s) \equiv \bm{n}(0),\, s\in[0,
  t]$, for any smooth function $\phi\in
C^{\infty}(\mathbb{R}^{|\bm{n}|_1}), |n|_1\coloneqq \sum_{i=1}^k n_i$,
we define the measure
\begin{equation}
\begin{aligned}
  \gamma^m(\phi, t) \coloneqq  \int_{\mathcal{C}^{|\bm{n}|_1}}\!\phi(\bm{X}_{\bm{n}}(t; \omega))
        S\big(t; \bm{X}(t, \omega)_{\bm{n}(t)}\big)
%
%
        \dd{m(\omega)},\quad
        \bm{X}(0; \omega)_{\bm{n}} = \bm{X}(0)_{\bm{n}(0)},
\end{aligned}
\end{equation}
where $\mathcal{C}^d\coloneqq\mathcal{C}([0, t], \mathbb{R}^d)$ (the
integration is taken all realizations of
$\bm{X}(t;\omega)_{\bm{n}}$). Using Theorem 6.2 in
\cite{lecavil2015probabilistic}, $\gamma^m(\phi)$ solves the PDE
\begin{equation}
  \f{\p\gamma^m}{\p t} + \sum_{i=1}^k\sum_{j=1}^{n_i}
  \f{\p(g_{i, j}(X_{i, j}(t), t)\gamma^m)}{\p X_{i, j}(t)}
  =  \mfrac{1}{2}\sum_{i=1}^k
  \sum_{j=1}^{n_i} \f{\p^2 (\sigma_{i, j}^2(X_{i, j}(t), t)\gamma^m)}{(\p X_{i, j}(t))^2}
  -\sum_{i=1}^k\sum_{j=1}^{n_i}\!\big(\beta_{i, j}(X_{i, j}(t))+
  \mu_{i, j}(X_{i, j}(t))\big) \gamma^m 
\label{gammapde}
\end{equation}
in the sense of distributions. Let $K^{\epsilon} =
\frac{1}{\epsilon^{|\bm{n}|_1}}K(\cdot)$, where $K(\cdot)$ is a smooth
mollifier,  and define

\begin{equation}
v^{\epsilon}(\bm{X}_{\bm{n}}, t) 
\coloneqq \gamma^m\big(K^{\epsilon}(\cdot - \bm{X}_{\bm{n}}), t\big),
\end{equation}
or, 

\begin{equation}
v^{\epsilon}(\bm{X}_{\bm{n}}, t) = \mathbb{E}
\Big[K^{\epsilon}\big(\bm{X}(t)_{\bm{n}(t)}-\bm{X}_{\bm{n}}\big)
S\big(t;\bm{X}(t)_{\bm{n}(t)}\big) \Big|
  \bm{X}(0)_{\bm{n}(0)}, 0; \bm{n}(0\leq \tau\leq t) = \bm{n}(0)\Big],
\end{equation}
where $S(t;\bm{X}(t)_{\bm{n}(t)})$ is the survival probability
defined in Eq.~\eqref{Wdef}.  By Eq.~\eqref{gammapde}, we have

\begin{equation}
  {\small
    \begin{aligned}
\f{\p v^{\epsilon}(\bm{X}_{\bm{n}}, t)}{\p t} = & \mathbb{E}
\Big[\sum_{i=1}^k\sum_{j=1}^{n_i}\partial_{X_{i, j}(t)} 
K^{\epsilon}\big(\bm{X}(t)_{\bm{n}(t)}\!-\!\bm{X}_{\bm{n}}\big) g_{i, j}(X_{i, j}(t), t) 
S\big(t; \bm{X}(t)_{\bm{n}(t)}\big)\Big|
\bm{X}(0)_{\bm{n}(0)}, 0; \bm{n}(0\leq \tau\leq t) = \bm{n}(0)\Big]\\
\: & +\mathbb{E}\Big[ \sum_{i=1}^k\sum_{j=1}^{n_i}\mfrac{1}{2}
\p_{X_{i, j}(t)}^2 K^{\epsilon}\big(\bm{X}(t)_{\bm{n}(t)}\!-\!\bm{X}_{\bm{n}}\big)
\sigma_{i, j}^2\big(X_{i, j}(t), t\big) S\big(t; \bm{X}(t)_{\bm{n}(t)}\big)
\Big|\bm{X}(0)_{\bm{n}(0)}, 0; \bm{n}(0\leq \tau\leq t) = \bm{n}(0)\Big]\\
\: & -\mathbb{E}\Big[\sum_{i=1}^k\sum_{j=1}^{n_i}
  \big(\beta_{i, j}(X_{i, j}(t))+\mu_{i, j}(X_{i, j}(t))\big)
  K^{\epsilon}(\bm{X}(t)_{\bm{n}(t)}-\bm{X}_{\bm{n}})
 S\big(t; \bm{X}(t)_{\bm{n}(t)}\big)
  \Big|\bm{X}(0)_{\bm{n}(0)}, 0; \bm{n}(0\leq \tau\leq t) = \bm{n}(0)\Big].
\end{aligned}}
\end{equation}

The assumptions that we shall impose for Proposition~\ref{lemma1} are
that: i) the limit

\begin{equation}
v\coloneqq\lim\limits_{\epsilon\rightarrow0^+}v^{\epsilon}=\mathbb{E}\Big[
\delta(\bm{X}(t)_{\bm{n}(t)}\!-\!\bm{X}_{\bm{n}})
S\big(t; \bm{X}(t)_{\bm{n}(t)}\big)\Big|
\bm{X}(0)_{\bm{n}(0)}, 0; \bm{n}(0\leq \tau\leq t) = \bm{n}(0)\Big]
\label{udef}
\end{equation}
exists, and ii) taking the limit $\epsilon\rightarrow 0^+$
commutes with taking the expectation and the derivative
w.r.t. $t$ and $X_{i, j}$, \textit{i.e.},

\begin{equation}
  {\small
    \begin{aligned}
\f{\p v(\bm{X}_{\bm{n}(t)}, t)}{\p t} = &
\mathbb{E}\Big[\sum_{i=1}^k\sum_{j=1}^{n_i}\partial_{X_{i, j}(t)} 
\delta\big(\bm{X}(t)_{\bm{n}(t)}\!-\!\bm{X}_{\bm{n}}\big) g_{i, j}\big(X_{i, j}(t), t\big)
S\big(t; \bm{X}(t)_{\bm{n}(t)}\big)\Big|\bm{X}(0)_{\bm{n}(0)}, 0; 
\bm{n}(0\leq \tau\leq t) = \bm{n}(0)\Big]\\
\: & +\mathbb{E}\Big[\sum_{i=1}^k\sum_{j=1}^{n_i}\mfrac{1}{2}\p_{X_{i, j}(t)}^2 
\delta\big(\bm{X}(t)_{\bm{n}(t)}\!-\!\bm{X}_{\bm{n}}\big) 
\sigma_{i, j}^2\big(X_{i, j}(t), t\big) 
S\big(t; \bm{X}(t)_{\bm{n}(t)}\big)\Big|
\bm{X}(0)_{\bm{n}(0)}, 0; \bm{n}(0\leq \tau\leq t) = \bm{n}(0)\Big]\\
\: & -\mathbb{E}\Big[\sum_{i=1}^k\sum_{j=1}^{n_i}\big(\beta_{i, j}(X_{i, j}(t))
+\mu_{i, j}(X_{i, j}(t))\big)\delta\big(\bm{X}(t)_{\bm{n}(t)}\!-\!\bm{X}_{\bm{n}}\big)
S\big(t; \bm{X}(t)_{\bm{n}(t)}\big)\Big|
\bm{X}(0)_{\bm{n}(0)}, 0; \bm{n}(0\leq \tau\leq t) = \bm{n}(0)\Big].
\end{aligned}}
\end{equation}
After integration by parts and noticing that

\begin{equation}
\begin{aligned}
g_{i, j}(X_{i, j}, t)v\equiv & \mathbb{E}
\Big[\sum_{i=1}^k\sum_{j=1}^{n_i} \delta(\bm{X}(t)_{\bm{n}(t)}-\bm{X}_{\bm{n}}) 
g_{i, j}(X_{i, j}(t), t)S\big(t; \bm{X}(t)_{\bm{n}(t)}\big)\Big|
\bm{X}(0)_{\bm{n}(0)}, 0; \bm{n}(0\leq \tau\leq t) = \bm{n}(0)\Big],\\
\sigma^2_{i, j}(X_{i, j}, t)v\equiv & \mathbb{E}
\Big[\sum_{i=1}^k\sum_{j=1}^{n_i} \delta(\bm{X}(t)_{\bm{n}(t)}\!-\!\bm{X}_{\bm{n}})
\sigma^2_{i, j}(X_{i, j}(t), t)S\big(t; \bm{X}(t)_{\bm{n}(t)}\big)\Big|
\bm{X}(0)_{\bm{n}(0)}, 0; \bm{n}(0\leq \tau\leq t) = \bm{n}(0)\Big],
\end{aligned}
\end{equation}
the partial differential equation satisfied by $v$ is

\begin{equation}
\f{\p v}{\p t} + \sum_{i=1}^k\sum_{j=1}^{n_i}\f{\p(g_{i}(X_{i, j}, t)v)}{\p X_{i, j}} 
= \mfrac{1}{2} \sum_{i=1}^k\sum_{j=1}^{n_i} \f{\p^2 (\sigma_{i, j}^2(X_{i, j}, t) v)}{(\p X_{i, j})^2}
-\sum_{i=1}^k\sum_{j=1}^{n_i}\big(\beta(X_{i, j})+\mu(X_{i, j})\big) v ,
\end{equation}
which proves Proposition~\ref{lemma1}.


\subsection{Proof of Proposition \ref{Markovjump}}
\label{appendixB}
We prove Proposition~\ref{lemma1} by induction on $m$. 
%
%
%
%
Clearly, when $m=0, 1$, $p^{0}$ and $p^{1}$ solve Eq.~\eqref{induct}
by using Proposition~\ref{lemma1}.  If the conclusion holds for
$m \geq 1$, then when $\bm{n}\neq \bm{n}(0)$, we have

\begin{equation}
  {\footnotesize
    \begin{aligned}
\f{\p p_{\bm{n}}^{m+1}}{\p t} = &
\mathbb{E}\Big[S\big(t; \bm{X}(t)_{\bm{n}(t)}\big)J^m(t,t; \bm{X}_{\bm{n}},\bm{n}(0))
  \Big|\bm{X}_{\bm{n}(0)}(0), 0; \bm{n}(0<s<t)=\bm{n}(0)
  \Big] \\
\: & +\int_0^t \mathbb{E}\Big[S\big(\tau; \bm{X}_{\bm{n}}\big)\partial_{t}
  J^m\big( t,\tau; \bm{X}_{\bm{n}},\bm{n}(0)\big) 
 \Big|\bm{X}(0)_{\bm{n}(0)}, 0; \bm{n}(0<s<\tau)=\bm{n}(0)\Big]\, \dd \tau\\
= & - \sum_{i=1}^k\sum_{j=1}^{n_i}\f{\p(g_{i, j}(X_{i, j}, t)p^{m+1}_{\bm{n}})}
   {\p X_{i, j}}+\mfrac{1}{2} \sum_{i=1}^k\sum_{j=1}^{n_i}
   \f{\p^2 (\sigma^2_{i, j}(X_{i, j}, t) 
  p^{m+1}_{\bm{n}})}{(\p X_{i, j})^2}
-\sum_{i=1}^k\sum_{j=1}^{n_i}\big(\beta_{i, j}(X_{i, j})
+\mu_{i, j}(X_{i, j})\big) p^{m+1}_{\bm{n}} \\
\: & +\sum_{i=1}^{k-1}\sum_{j=1}^{n_i+1}\!\!\int\!
\tilde{\beta}_{i, j}(Y, X_{i+1, n_{r+1}-1}, X_{r+1, n_{i+1}})
\mathbb{E}\Big[\delta({X}_{i, j}(t)-Y)
  \delta(\bm{X}(t)_{{\bm{n}(0)_{\text{b}, -i}}}^{{{-j}}}\!\!\!\!-\!\bm{X}_{{\bm{n}}}) \\[-4pt]
\: & \hspace{4.5cm}\times
\delta_{{\bm{n}(0)_{\text{b}, -i}}, \bm{n}} S\big(t; \bm{X}_{\bm{n}}\big)
\Big|\bm{X}(0)_{{\bm{n}(0)}}, 0;\bm{n}(0<s<t)=\bm{n}(0)\Big]\dd {Y}\\
\: & + \sum_{i=1}^{k-1}\sum_{j=1}^{n_i+1}\int
\tilde{\beta}_{i, j}(Y, X_{i+1, n_{i+1}-1}, X_{i+1, n_{i+1}}) \\[-4pt]
\: &  \hspace{3cm}  \times \bigg(\int_0^t \mathbb{E}
\Big[S\big(\tau; \bm{X}_{\bm{n}}\big)
  J^{m-1}\big( t,\tau;\bm{X}_{{\bm{n}_{\text{b}, i}}}^{{{j}}},\bm{n}(0)\big)
 \Big|\bm{X}(0)_{\bm{n}(0)}, 0;\bm{n}(0<s<t)=\bm{n}(0)
\Big]  \dd{\tau}\bigg)\, \dd{Y}\\
\: & + \sum_{i=1}^{\infty}\sum_{j=1}^{n_i+1}\int \mu_{i, j}(Y)
\mathbb{E}\Big[\delta(X_{i, j}-Y)
  \delta(\bm{X}_{\bm{n}(0)_{\text{d}, -i}}^{{ -j}}(t)\!-\!\bm{X}_{\bm{n}})
  \delta_{\bm{n}(0)_{\text{d}, -i}, \bm{n}}
  S\big(t;\bm{X}(t)_{\bm{n}(t)}\big)\Big|\bm{X}(0)_{\bm{n}(0)},
  0;\bm{n}(0<s<t)=\bm{n}(0) \Big]\dd{Y} \\
\: & + \sum_{i=1}^{\infty}\sum_{j=1}^{n_i+1}\int \mu_{i, j}(Y)
\int_0^t\mathbb{E}\Big[S\big(\tau; \bm{X}_{\bm{n}}\big)
  J^{m-1}\big(t,\tau;\bm{X}_{\bm{n}_{\text{d}, i}}^{{ -j}},\bm{n}(0)\big)
  \Big|\bm{X}(0)_{\bm{n}(0)}, 0\Big] \dd{\tau}\, \dd{Y}\\
= & - \sum_{i=1}^k\sum_{j=1}^{n_i}\f{\p(g_{i, j}p^{m+1}_{\bm{n}})}{\p X_{i, j}}
+ \mfrac{1}{2} \sum_{i=1}^k\sum_{j=1}^{n_i} \f{\p^2 (\sigma_{i, j}
  p^{m+1}_{\bm{n}})}{(\p X_{i, j})^2}
-\sum_{i=1}^k\sum_{j=1}^{n_i}\big(\beta_{i, j}(X_{i, j})
+\mu_{i, j}(X_{i, j})\big) p^{m+1}_{\bm{n}} \\
\: & +\sum_{i=1}^{k-1}\sum_{j=1}^{n_i+1}\int
\tilde{\beta}(Y, X_{i+1, n_{i+1}-1}, X_{i+1, n_{i+1}})
p^{m}_{\bm{n}_{\text{b}, i}}(\bm{X}_{\bm{n}_{\text{b}, i}}^{{ j}}, t\vert
\bm{X}(0)_{\bm{n}(0)}, 0)\dd{Y}\\
\: & + \sum_{i=1}^{\infty}\sum_{j=1}^{n_i+1}\int
\mu(Y)p^{m}_{\bm{n}_{\text{d}, i}}(\bm{X}_{\bm{n}_{\text{d}, i}}^{{j}}, t|
\bm{X}(0)_{\bm{n}(0)}, 0) \dd{Y}.
\end{aligned}}
\label{induct_step2}
\end{equation}

Here, the function $\delta_{\bm{n}(0)_{\text{b}, -i}, \bm{n}}=1$ if
$\bm{n}(0)_{\text{b}, -i}=\bm{n}$ and $\delta_{\bm{n}(0)_{\text{b},
    -i}, \bm{n}}=0$ otherwise; similarly, $\delta_{\bm{n}(0)_{{\rm d},
    -i}, \bm{n}}=1$ if $\bm{n}(0)_{{\rm d}, -i}=\bm{n}$ and
$\delta_{\bm{n}(0)_{{\rm d}, -i}, \bm{n}}=0$
otherwise. Proposition~\ref{lemma1} shows that

\begin{equation}
  \mathbb{E}\big[\delta(\bm{X}(t)_{\bm{n}}\!-\!\bm{X}_{\bm{n}})
    S(t; \bm{X}(t)_{\bm{n}(t)})
    \big|\bm{X}(0)_{\bm{n}(0)}, 0; \bm{n}(0\leq s \leq t) = \bm{n}(0)\big]
\end{equation}
satisfies Eq.~\eqref{forwardFeymann}, so we can verify that
Eq.~\eqref{induct} also holds for $m+1$ when $\bm{n}=
\bm{n}(0)$. Thus, we have proved that Eq.~\eqref{induct} holds true
for $m+1$.  Additionally, it is obvious that $p^{m+1}_{\bm{n}} \geq
p^{m}_{\bm{n}}$ holds for $m=0$. If $p^{m}_{\bm{n}} \geq
p^{m-1}_{\bm{n}}$ for any $\bm{n}, \bm{X}_{\bm{n}}$, we have for
$\Delta^{m}_{\bm{n}} \coloneqq p^{m}_{\bm{n}} - p^{m-1}_{\bm{n}}$,

\begin{equation}
\begin{aligned}
  \Delta^{m+1}_{\bm{n}} = \int_0^t\mathbb{E}\Big[S\big(\tau; & \bm{X}_{\bm{n}}\big) 
    \sum_{i=1}^{k(0)}\sum_{j=1}^{n_i(0)}
    \Big(\tilde{\beta}_{i, j}(X_{i, j}(\tau),X_1, X_2)
    \Delta_{\bm{n}}^{m}(\bm{X}_{\bm{n}}, t-\tau|\bm{X}(\tau)_{\bm{n}(0)_{\text{b}, -i}}^{{-j}}, 0) \\
    \: & \,\, + \mu_{i, j}(X_{i, j})
    \Delta^{m}_{\bm{n}}(\bm{X}_{\bm{n}}, t-\tau|\bm{X}(\tau)_{\bm{n}_{\text{d}, -i}}^{{-j}}, 0)
    \Big)\Big| \bm{X}(0)_{\bm{n}(0)}, 0; \bm{n}(0\leq s \leq \tau)
    = \bm{n}(0)\Big]\,\dd{\tau} \geq 0.
\end{aligned}
\end{equation}
Therefore, we have proved that $p^{m+1}_{\bm{n}}$ satisfies
Eq.~\eqref{induct} and that $p^{m+1}_{\bm{n}} \geq p^{m}_{\bm{n}}$ for
all $m\in\mathbb{N}$ by induction.

\section{Differential equations satisfied by $X^q(t), q\in\mathbb{N}^+$}
\label{appendixc}
\setcounter{equation}{0}
\renewcommand{\theequation}{B\arabic{equation}}

\noindent With $X^q(t)$ according to Eq.~\eqref{Xhigher}, it can be
  shown that for $q> 1$,

\begin{equation}
\begin{aligned}
  \frac{\dd X^q(t) }{\dd t} = & q\sum_{\bm{n}}\int
  \Big(\sum_{i=1}^{k}\sum_{j=1}^{n_i}X_{i, j}\Big)^{q-1}
  \Big(\sum_{\ell=1}^{k}
  \sum_{m=1}^{n_{\ell}} g_{\ell}(X_{\ell, m}, t)\Big)
  \rho_{\bm{n}}(\bm{X}_{\bm{n}}, t)\dd {\bm{X}_{\bm{n}}}\\
  \: & +\frac{q(q-1)}{2}\sum_{\bm{n}}\int
  \Big(\sum_{i=1}^{k}\sum_{j=1}^{n_i}X_{i, j}\Big)^{q-2}
        \Big(\sum_{\ell=1}^{k}\sum_{m=1}^{n_{\ell}}
  \sigma^2_{\ell}(X_{\ell, m}, t)\Big)
  \rho_{\bm{n}}(\bm{X}_{\bm{n}}, t)\dd {\bm{X}_{\bm{n}}}\\
  \: & - \sum_{\bm{n}}\int
  \bigg[\sum_{i=1}^{k}\sum_{j=1}^{n_i}\mu_i(X_{i, j}, t)
   \sum_{r=1}^{q}(-1)^{r-1}\binom{q}{r}
   \Big(\sum_{\ell=1}^{k}{\hspace{-1pt}\vphantom{\sum}}'
   \sum_{m=1}^{n_{\ell}}{\hspace{-2pt}\vphantom{\sum}}'X_{\ell, m}\Big)^{q-r} X_{i, j}^{r}\bigg]
    \rho_{\bm{n}}(\bm{X}_{\bm{n}}, t)\dd {\bm{X}_{\bm{n}}} \\
    \: & -\sum_{\bm{n}}\int
   \Big(\sum_{i=1}^{k}\sum_{j=1}^{n_i}X_{i, j}\Big)^q
      \sum_{\ell=1}^{k}\sum_{m=1}^{n_{\ell}}\beta_{\ell}(X_{\ell, m})
    \rho_{\bm{n}}(\bm{X}_{\bm{n}}, t) \dd {\bm{X}_{\bm{n}}}\\
    \: & +\sum_{\bm{n}} \int\Big(\sum_{i=1}^{k}\sum_{j=1}^{n_i} X_{i, j}\Big)^q
 \bigg(\sum_{\ell=1}^{k-1}\frac{n_{\ell}+1}{n_{\ell+1}(n_{\ell+1}-1)}
 \medint\int  \medop\sum_{j_1\neq j_2}\tilde{\beta}(Y, X_{\ell+1, j_1}, X_{\ell+1, j_2})
    \rho_{\bm{n}_{\text{b}, \ell}}(\bm{X}_{\bm{n}_{\text{b}, \ell}}^{{ j_1, j_2}}, t)
    \dd{Y}\bigg) \dd {\bm{X}_{\bm{n}}},
\end{aligned}
\end{equation}
where $\rho_{\bm{n}}$ is the symmetric probability density function
defined in Eq.~\eqref{rhodef}. Here,
$\textstyle\sum\limits_{\ell=1}^{k}{}^{\!'}\textstyle\sum\limits_{m=1}^{n_{\ell}}{}^{\!\!'}$
denote sums over which $\ell \neq i$ or $m\neq j$.

In particular, if $X$ is a conserved quantity at division, then the
evolution of the second-order moment can be further simplified as

\begin{equation}
\begin{aligned}
  \frac{\dd X^q(t)}{\dd t} = & q\sum_{\bm{n}}\int
       \Big(\sum_{i=1}^{k}\sum_{j=1}^{n_i}X_{i, j}\Big)^{q-1}
         \sum_{\ell=1}^{k}\sum_{m=1}^{n_{\ell}}
         g_{\ell}(X_{\ell, m}, t)
    \rho_{\bm{n}}(\bm{X}_{\bm{n}}, t)\dd {\bm{X}_{\bm{n}}}\\
  \: & +\sum_{\bm{n}}\mfrac{q(q-1)}{2}\int
  \Big(\sum_{i=1}^{k}\sum_{j=1}^{n_i}X_{i, j}\Big)^{q-2}
    \sum_{\ell=1}^{k}\sum_{m=1}^{n_{\ell}}
    \sigma^2_{\ell}\big(X_{\ell, m}, t\big)
  \rho_{\bm{n}}(\bm{X}_{\bm{n}}, t)\dd {\bm{X}_{\bm{n}}} \\
  \: & - \sum_{\bm{n}}\int \Big[\sum_{i=1}^{k}\sum_{j=1}^{n_i}\mu_i(X_{i, j})
   \sum_{r=1}^{q}(-1)^{r-1}\binom{q}{r}
      \Big(\sum_{\ell=1}^{k}{\hspace{-1pt}\vphantom{\sum}}' 
    \sum_{m=1}^{n_{\ell}}{\hspace{-2pt}\vphantom{\sum}}' X_{\ell, m}\Big)^{q-r} X_{i, j}^{r} \Big]
  \rho_{\bm{n}}(\bm{X}_{\bm{n}}, t)\dd {\bm{X}_{\bm{n}}}.
\end{aligned}
\label{Xsquare}
\end{equation}
Eq.~\eqref{Xsquare} can be further simplified if the coefficients
$g_i$ and $\sigma_i$ satisfy certain conditions. For example, if the
cells grow exponentially, $\textit{i.e.}$, $g_i(X_{i, j}, t) = \lambda
X_{i, j}$ and $\sigma_i^2(X_{i, j}, t) = \sigma^2 X_{i, j}$.
Eq.~\eqref{Xsquare} can be more simply expressed as

\begin{equation}
\begin{aligned}
  \frac{\dd X^q(t) }{\dd t} = & \lambda q X^q(t) +
  \mfrac{q(q-1)}{2} \sigma^2 X^{q-1}(t)
\\
\: & - \sum_{\bm{n}}\int\Big[ \sum_{i=1}^{k}\sum_{j=1}^{n_i}\mu_i(X_{i, j})
\sum_{r=1}^{q}(-1)^{i-1}\binom{q}{r}
\Big(\sum_{\ell=1}^{k}{\hspace{-1pt}\vphantom{\sum}}'
\sum_{m=1}^{n_{\ell}}{\hspace{-2pt}\vphantom{\sum}}' X_{\ell, m}\Big)^{q-r}
X_{i, j}^{r}\Big]\rho_{\bm{n}}(\bm{X}_{\bm{n}}, t)
\dd {\bm{X}_{\bm{n}}}.
\end{aligned}
\end{equation}

\section{Birth-induced boundary conditions}
\label{appendixd}
\setcounter{equation}{0}
\renewcommand{\theequation}{C\arabic{equation}}

We can also consider variables that describe cellular quantities that
reset upon cell division. Example of such variables include cell size
and cell age \cite{Xia2020, Xia_2021}.  Specifically, consider simple
``timer'' models where a new daughter cell acquires age 0 at its
birth, while the other cell is assumed to be the ``mother'' that
continues to age. This assignment of age across a proliferating
population is described as ``budding'' birth
\cite{chou2016hierarchical_PRE,chou2016hierarchical}.  \added{A
  kinetic theory can track both cell volume and cell age through the
  variables $\bm{X}_{\bm{n}}$ and $\bm{A}_{\bm{n}}\coloneqq
  (\bm{A}_1,...,\bm{A}_k)$, respectively. Here, in analogy with
  $X_{i,j} \, (j \leq n_{i})$ (Table~\ref{tab:model_variables}),
  $\bm{A}_i\coloneqq (A_{i,1},...,A_{i, n_i})$ and $A_{i, j}\, (j\leq
  n_{i})$ is the age of the $j^{\rm th}$ cell of generation $i$.}

We can show that the solution to
\begin{equation}
\begin{aligned}
\f{\p \hat{p}_{\bm{n}}(\bm{A}_{\bm{n}}, \bm{X}_{\bm{n}}, t)}{\p t}
  + & \sum_{i=1}^k\sum_{j=1}^{n_i}\f{\p\hat{p}_{\bm{n}}}{\p A_{i, j}}
  + \sum_{i=1}^k\sum_{j=1}^{n_i}\f{\p(g_{i, j}(A_{i, j},X_{i, j}, t)
    \hat{p}_{\bm{n}})}{\p X_{i, j}} \\
  \: &  =  \mfrac{1}{2} \sum_{i=1}^k\sum_{j=1}^{n_i}
  \f{\p^2 (\sigma_{i, j}^2(A_{i, j},X_{i, j}, t)
   \hat{p}_{\bm{n}})}{(\p X_{i, j})^2} -\sum_{i=1}^k\sum_{j=1}^{n_i}\big(\beta_{i, j}(A_{i, j},X_{i,
   j})+\mu_{i, j}(A_{i, j},X_{i, j})\big) \hat{p}_{\bm{n}}\\
 \hat{p}_{\bm{n}}(\bm{A}_{\bm{n}}, \bm{X}_{\bm{n}},
 0| \bm{A}(0)_{\bm{n}(0)},&\, \bm{X}(0)_{\bm{n}(0)}, 0) =  \delta(\bm{X}(0)_{\bm{n}(0)} -
 \bm{X}_{\bm{n}})\delta(\bm{A}(0)_{\bm{n}(0)} - \bm{A}_{\bm{n}}),\quad 
 ~\text{if}~\bm{n} = \bm{n}(0), \\[6pt]
\: &\! \hat{p}_{\bm{n}}(\bm{A}_{\bm{n}}, \bm{X}_{\bm{n}}, 0)= 0 ~\qquad
 \text{if}~\bm{n} \neq \bm{n}(0)
\end{aligned}
\label{forwardFeymann1}
\end{equation}
can be expressed as 

\begin{equation}
\begin{aligned}
  \hat{p}_{\bm{n}}(\bm{A}_{\bm{n}}, \bm{X}_{\bm{n}}, t|\bm{A}(0)_{\bm{n}(0)},
  \bm{X}(0)_{\bm{n}(0)}, 0) \coloneqq &
  \mathbb{E}\Big[\delta(\bm{X}(t)_{\bm{n}(t)}\!-\!\bm{X}_{\bm{n}})
    \delta(\bm{A}(t)_{\bm{n}(t)}\!-\!\bm{A}_{\bm{n}}))\added{S_A\big(t; \bm{X}(t)_{\bm{n}(t)}, \bm{A}(t)_{\bm{n}(t)}\big)}
    \\
    \: &\qquad 
      \Big| \bm{A}(0)_{\bm{n}(0)},
      \bm{X}(0)_{\bm{n}(0)}, 0;\bm{n}(0\leq s \leq t) = \bm{n}(0)\Big],\\
  \hat{p}_{\bm{n}}(\bm{A}_{\bm{n}}, \bm{X}_{\bm{n}}, t) = & 0,
  \qquad  ~\text{if}~\bm{n} \neq \bm{n}(0),
\end{aligned}
\end{equation}
where here,

\begin{equation}
\begin{aligned}
  S_A\big(\added{t;\bm{X}(t)_{\bm{n}(t)}, \bm{A}(t)_{\bm{n}(t)}}\big)\coloneqq \exp
  \Big(-\int_{0}^{t}\sum_{i=1}^{k(0)}\sum_{j=1}^{n_{i}(0)}
  \big(\beta(A_{i,j}(\tau), X_{i, j}(\tau))+\mu(A_{i,j}(\tau), X_{i, j}(\tau))\big)
  \dd \tau\Big).
\end{aligned}
\label{Sdef}
\end{equation}
Furthermore, if we set
\begin{equation}
\begin{aligned}
  p^{0}(\bm{A}_{\bm{n}}, \bm{X}_{\bm{n}}, t|\bm{A}(0)_{\bm{n}(0)},
  \bm{X}(0)_{\bm{n}(0)}, 0) = & 0,\\
  p^{1}(\bm{A}_{\bm{n}}, \bm{X}_{\bm{n}}, t|\bm{A}(0)_{\bm{n}(0)},
  \bm{X}(0)_{\bm{n}(0)}, 0) = & \hat{p}_{\bm{n}}(\bm{A}_{\bm{n}},
  \bm{X}_{\bm{n}}, t|\bm{A}(0)_{\bm{n}(0)}, \bm{X}(0)_{\bm{n}(0)}, 0)
\end{aligned}
\end{equation}
we can define the recursion
\begin{equation}
  {\small
    \begin{aligned}
& p_{\bm{n}}^{m+1}(\bm{A}_{\bm{n}}, \bm{X}_{\bm{n}}, t
  |\bm{A}(0)_{\bm{n}(0)}, \bm{X}(0)_{\bm{n}(0)}, 0)
  = \hat{p}_{\bm{n}}(\bm{A}_{\bm{n}}, \bm{X}_{\bm{n}}, t|
  \bm{A}(0)_{\bm{n}(0)}, \bm{X}(0)_{\bm{n}(0)}, 0) \\
  & \hspace{6mm}+\mathbb{E}\Big[\int_0^t \!S_A(\tau; \bm{X}(\tau)_{\bm{n}(\tau)},
    \bm{A}(\tau)_{\bm{n}(\tau)})
    J_{A}^{m}\big(t, \tau;\bm{X}_{\bm{n}},\bm{A}_{\bm{n}}, \bm{n}(0)\big)
    \dd{\tau}\Big|\bm{A}(0)_{\bm{n}(0)},
    \bm{X}(0)_{\bm{n}(0)}, 0;\bm{n}(0\leq \tau\leq t) = \bm{n}(0)\Big], \,\,
  ~\text{if}~ \bm{A}_{\bm{n}}>0, \\
& p_{\bm{n}}^{m+1}(\bm{A}_{\bm{n}}, \bm{X}_{\bm{n}}, t
  |\bm{A}(0)_{\bm{n}(0)}, \bm{X}(0)_{\bm{n}(0)}, 0) =\\
& \hspace{1.3cm} \mathbb{E}\Big[\sum_{j=1}^{{n_{i}}}
    \tilde{\beta}(A_{i, j}, Y(t),X_{i, j}(t), X_{i+1, n_{i+1}})
    p^{m}_{\bm{n}}(\bm{A}_{{\bm{n}_{\text{b}, i}}}^{{{j}}},
    \bm{X}_{{\bm{n}_{\text{b}, i}}}^{{{j}}}, t|\bm{A}(0)_{\bm{n}(0)},
    \bm{X}(0)_{\bm{n}(0)},0)\Big], \qquad\quad \text{if}~ A_{i+1, n_{i+1}} = 0,\\
& p_{\bm{n}}^{m+1}(\bm{A}_{\bm{n}}, \bm{X}_{\bm{n}}, t
|\bm{A}(0)_{\bm{n}(0)}, \bm{X}(0)_{\bm{n}(0)}, 0) = 0,\qquad\qquad\text{otherwise}.
\end{aligned}}
\label{recursion_new}
\end{equation}

Here, $\bm{A}_{\bm{n}}>0$ indicates that each component in
$\bm{A}_{i}$ of $\bm{A}_{\bm{n}}$ is greater than
0. $\tilde{\beta}(A_{i, j}, Y(t), X_{i, j}, X_{i+1, n_{i+1}})$ is the
rate of a cell in the $i^{\text{th}}$ generation giving birth to a
cell in the $(i+1)^{\text{th}}$ generation with the state $X_{i+1,
  n_{i+1}}$ and its own state shifting to $X_{i,
  j}$. $\bm{A}_{{\bm{n}_{\text{b}, i}}}^{{{j}}}$ differs from
$\bm{A}_{{\bm{n}}}$ in that its $(i+1)^{\text{th}}$ generation does
not contain the $(n_{i+1})^{\text{th}}$
component. $\bm{X}_{{\bm{n}_{\text{b}, i}}}^{{{j}}}$ differs from
$\bm{X}_{{\bm{n}}}$ in that its $j^{\text{th}}$ component in the
$i^{\text{th}}$ generation is $Y_{i, j}(t)$, not $X_{i,j}$ and it does
not have the $(n_{i+1})^{\text{th}}$ component in the
$(i+1)^{\text{th}}$ generation. In analogy to Eq.~\eqref{Jm}, $J_A(t,
\tau; \bm{A}_{\bm{n}}, \bm{X}_{\bm{n}})$ in Eq.~\eqref{recursion_new}
is defined as

\begin{equation}
  {\small
    \begin{aligned}
J_{A}^{m}\big(t, \tau;  \bm{X}_{\bm{n}}, \bm{A}_{\bm{n}}, \bm{n}(0)\big) \coloneqq &
\sum_{i=1}^{k(0)}\sum_{j=1}^{n_{i}(0)}\!\Big[\tilde{\beta}_{i, j}\big(A_{i, j}(\tau),
  X_{i, j}(\tau), X_1(\tau), X_2(\tau)\big)
  p_{\bm{n}}^m(\bm{A}_{\bm{n}}, \bm{X}_{\bm{n}}, t-\tau|\bm{A}(\tau)_{\bm{n}(0)_{ \text{b}, -i}}^{-j},
  \bm{X}(\tau)_{\bm{n}(0)_{\text{b}, -i}}^{-j}, 0) \\
  \: & \qquad \hspace{1.4cm}+ \mu_{i, j}(A_{i, j}(\tau), X_{i, j}(\tau))
  p_{\bm{n}}^m(\bm{A}_{\bm{n}}, \bm{X}_{\bm{n}}, t-\tau|\bm{A}(\tau)_{\bm{n}(0)_{ \text{d}, -i}}^{-j}, 
\bm{X}(\tau)_{\bm{n}(0)_{ \text{d}, -i}}^{-j}, 0)\Big].
\end{aligned}}
\label{JAdef}
\end{equation}
In Eq.~\eqref{JAdef}, $\bm{A}_{{\bm{n}(0)_{\text{b}, -i}}}^{{{-j}}}$
differs from $\bm{A}_{{\bm{n}({0})}}$ in that its $(i+1)^{\text{th}}$
generation has an extra component $A_{i+1, n_{i+1}(0)+1}
=0$. $\bm{X}_{{\bm{n}({0})_{\text{b}, -i}}}^{{{-j}}}(\tau)$ is
different from $\bm{X}_{{\bm{n}({0})}}(\tau)$ in that compared to
$\bm{X}_{{\bm{n}({0})}}(\tau)$, the $j^{\text{th}}$ component of the
$i^{\text{th}}$ generation of $\bm{X}_{{\bm{n}({0})_{\text{b},
      -i}}}^{{{-j}}}(\tau)$ is $Y_{i, j}(\tau)$ in the
$j^{\text{th}}$, but the $j^{\text{th}}$ component of the
$i^{\text{th}}$ generation is $X_{i, j}(\tau)$ for
$\bm{X}_{\bm{n}(0)}(\tau)$; furthermore, the $(i+1)^{\text{th}}$
generation of $\bm{X}_{{\bm{n}({0})}}(\tau)$ does not have the
$(n_{i+1}+1)^{\text{th}}$ component $X_{i+1,
  n_{i+1}(0)+1}(\tau)$. $\bm{A}(\tau)_{{\bm{n}(0)_{\text{d},
      -i}}}^{{{-j}}}$ differs from $\bm{A}(\tau)_{{\bm{n}}}$ in that
its $i^{\text{th}}$ generation is $(A_{i, 1}(\tau),...,A_{i,
  j-1}(\tau), A_{i, j+1}(\tau),...,A_{i, n_i(0)}(\tau))$, and
$\bm{X}(\tau)_{{\bm{n}(0)_{\text{d}, -i}}}^{{{-j}}}$ differs from
$\bm{X}(\tau)_{{\bm{n}}}$ in that its $i^{\text{th}}$ generation is
$(X_{i, 1}(\tau),...,X_{i, j-1}(\tau), X_{i, j+1}(\tau),...,X_{i,
  n_i(0)}(\tau))$.

Then, similar to the proof of Proposition~\ref{Markovjump},
$p_{\bm{n}}^{m+1}$ satisfies the following PDE

\begin{equation}
\begin{aligned}
  \f{\p p^{m+1}_{\bm{n}}(\bm{A}_{\bm{n}}, \bm{X}_{\bm{n}}, t)}{\p t}
  + & \sum_{i=1}^k\sum_{j=1}^{n_i}\f{\p p^{m+1}_{\bm{n}}}{\p A_{i, j}}
  + \sum_{i=1}^k\sum_{j=1}^{n_i}\f{\p(g_{i, j}(A_{i, j},X_{i, j}, t)
    p^{m+1}_{\bm{n}})}{\p X_{i, j}} \\
  = & \mfrac{1}{2}
  \sum_{i=1}^k\sum_{j=1}^{n_i} \f{\p^2 (\sigma_{i, j}^2(A_{i, j},X_{i, j}, t)
    p^{m+1}_{\bm{n}})}{(\p X_{i, j})^2}
  -\sum_{i=1}^k\sum_{j=1}^{n_i}\big(\beta_{i, j}(A_{i, j},X_{i, j})
  +\mu_{i, j}(A_{i, j},X_{i, j})\big) p^{m+1}_{\bm{n}}\\
  \: & + \sum_{i=1}^{\infty}\sum_{j=1}^{{n_i^{\text{d}}}}\int\mu(A, Y)
  p^{m}_{{\bm{n}_{\text{d}, i}}}(\bm{A}_{\bm{n}_{\text{d}, i}}^j,
  \bm{X}_{\bm{n}_{\text{d}, i}}^{{ j}}, t|\bm{A}(0)_{\bm{n}(0)},
  \bm{X}(0)_{\bm{n}(0)}, 0) \dd{Y}\dd{A}~,\qquad \text{if}~ \bm{A}_{\bm{n}}>0\\[8pt]
  p_{\bm{n}}^{m+1}(\bm{X}_{\bm{n}},\bm{A}_{\bm{n}}, t &
  |\bm{X}(0)_{\bm{n}(0)}, \bm{A}(0)_{\bm{n}(0)}, 0) \\
\: &  \hspace{-1cm}= \int \sum_{i=1}^{k-1}\sum_{j=1}^{{n_{i}}}
  \tilde{\beta}_{i, j}(A_{i, j}, Y_{i, j}, X_{i, j},X_{i+1, n_{i+1}})
  p^{m}_{\bm{n}}(\bm{A}_{{\bm{n}(0)_{\text{b}, i}}}^{j}(t),
  \bm{X}_{{\bm{n}(0)_{\text{b}, i}}}^{j}(t), t|
  \bm{A}(0)_{\bm{n}(0)},\bm{X}(0)_{\bm{n}(0)}, 0) \dd{Y_{i, j}},\\
 \:  &\qquad~\text{if}~ A_{i+1, n_{i+1}} = 0.
\end{aligned}
\end{equation}
%
%
Likewise, it can be shown that $p_{\bm{n}}^{m}$ is non-negative,
increasing in $m$, and satisfies

\begin{equation}
  \sum_{{\bm{n}}}\int p_{\bm{n}}^{m}(\bm{A}_{\bm{n}},
  \bm{X}_{\bm{n}}, t|\bm{A}(0)_{\bm{n}(0)},\bm{X}(0)_{\bm{n}(0)}, 0)
  \dd{\bm{X}_{\bm{n}}}\dd{\bm{A}_{\bm{n}}} \leq 1,
  \quad\forall \bm{A}(0)_{\bm{n}(0)},\bm{X}(0)_{\bm{n}(0)}.
\end{equation}
Therefore, under certain technical conditions
such as commuting derivatives, there exists a limit
$p_{\bm{n}}^{*} = \lim\limits_{m\rightarrow\infty} p_{\bm{n}}^{m}$
that satisfies the PDE

\begin{equation}
\begin{aligned}
  \f{\p p^{*}_{\bm{n}}(\bm{A}_{\bm{n}}, \bm{X}_{\bm{n}}, t)}{\p t}
  + & \sum_{i=1}^k\sum_{j=1}^{n_i}\f{\p p^{*}_{\bm{n}}}{\p A_{i, j}}
  +\sum_{i=1}^k\sum_{j=1}^{n_i}\f{\p(g_{i, j}(A_{i, j},X_{i, j}, t)p^{*}_{\bm{n}})}{\p X_{i, j}}\\
  = &  \mfrac{1}{2}
  \sum_{i=1}^k\sum_{j=1}^{n_i} \f{\p^2 (\sigma_{i, j}^2(A_{i, j},X_{i, j}, t)
    p^{*}_{\bm{n}})}{(\p X_{i, j})^2} -\sum_{i=1}^k\sum_{j=1}^{n_i}\big(\beta_{i, j}(A_{i, j}, X_{i, j})
  +\mu_{i, j}(A_{i, j}, X_{i, j})\big) p^{*}_{\bm{n}}\\
  \: &   + \sum_{i=1}^{\infty}\sum_{j=1}^{{n_i^{\text{d}}}}\int\mu_{i, j}(A, Y)
  p^{*}_{{\bm{n}_{\text{d}, i}}}(\bm{A}_{\bm{n}_{\text{d}, i}}^{{ j}},\bm{X}_{\bm{n}_{\text{d}, i}}^{{ j}}, t
  |\bm{A}(0)_{\bm{n}(0)},\bm{X}(0)_{\bm{n}(0)}, 0) \dd{Y}\dd{A}~, \quad
  \text{if}~ \bm{A}_{\bm{n}}>0, \\[8pt]
p_{\bm{n}}^{*}(\bm{A}_{\bm{n}},\bm{X}_{\bm{n}},t |\bm{X}(0)_{\bm{n}(0)},& \bm{A}(0)_{\bm{n}(0)},0) = \\
\int \sum_{i=1}^{k-1}\sum_{j=1}^{{n_{i}}} & 
  \tilde{\beta}_{i, j}(A_{i, j},Y_{i, j},X_{i, j},X_{i+1, n_{i+1}})
  p^{*}_{\bm{n}}\big(\bm{A}_{{\bm{n}(0)_{\text{b}, -i}}}^{{{-j}}},
  \bm{X}_{{\bm{n}(0)_{\text{b}, -i}}}^{-j}, t|
  \bm{A}(0)_{\bm{n}(0)},\bm{X}(0)_{\bm{n}(0)}, 0\big) \dd{Y_{i, j}},\\
  \: & \qquad  ~\text{if}~ A_{i+1, n_{i+1}} = 0.
\end{aligned}
\end{equation}
If $p^*$ satisfies the normalization conditions, $\textit{i.e.},
\sum_{\bm{n}}\int p^*_{\bm{n}}(\bm{A}_{\bm{n}},\bm{X}_{\bm{n}}, t)\dd
\bm{X}_{\bm{n}}\dd \bm{A}_{\bm{n}}\equiv 1, \forall t\geq 0$, we can
also define the unconditional probability density by averaging over
the initial probability density
$q_{\bm{n}(0)}\big(\bm{A}(0)_{\bm{n}(0)},\bm{X}(0)_{\bm{n}(0)},
0\big)$

\begin{equation}
\begin{aligned}
  p_{\bm{n}}\big(\bm{A}_{\bm{n}},\bm{X}_{\bm{n}}, t\big)\coloneqq \sum_{\bm{n}(0)}
  \int p_{\bm{n}}^*(\bm{A}_{\bm{n}}, \bm{X}_{\bm{n}},t|
  \bm{A}(0)_{\bm{n}(0)},\bm{X}(0)_{\bm{n}(0)}, 0)
  q_{\bm{n}(0)}\big(\bm{A}(0)_{\bm{n}(0)},
  \bm{X}_{\bm{n}(0)}, 0\big) \dd \bm{X}(0)_{\bm{n}(0)}\dd \bm{A}(0)_{\bm{n}(0)}.
\end{aligned}
\label{psdef_new}
\end{equation}

From Eq.~\eqref{psdef_new}, we can define the symmetric probability
density function
\begin{equation}
  \rho_{\bm{n}}(\bm{A}_{\bm{n}}, \bm{X}_{\bm{n}}, t) \coloneqq
  \prod_{i=1}^k \frac{1}{n_i!} \sum_{\pi} p_{\bm{n}}^{*}\big(\pi(\bm{A}_{\bm{n}}),
  \pi(\bm{X}_{\bm{n}}), t\big),
\label{rhodef_new}
\end{equation}
where $\pi$ is the same rearrangement for the age variables
$\bm{A}_{\bm{n}}$ and state variables $\bm{X}_{\bm{n}}$.  From
Eq.~\eqref{rhodef_new}, we could derive the macroscopic quantities
such as the marginalized cell density. We shall omit detailed
discussions on those macroscopic quantities for brevity.

\end{document}